\documentclass{article}
\usepackage[utf8]{inputenc}
\usepackage[english]{babel}
\usepackage{makecell}
\usepackage{longtable}
\usepackage{natbib}
\usepackage{graphicx}
\usepackage{booktabs}
\usepackage{threeparttablex}
\usepackage[dvipsnames]{xcolor}
\usepackage{amssymb}
\usepackage{pifont}
\newcommand{\xmark}{\ding{55}}%

\usepackage[width=1\textwidth]{caption}
\usepackage{rotating}
\usepackage{pdflscape}

\usepackage{arxiv}

\usepackage[utf8]{inputenc} 
\usepackage[T1]{fontenc}    
\usepackage{hyperref}       
\usepackage{url}            
\usepackage{booktabs}       
\usepackage{amsfonts}       
\usepackage{nicefrac}       
\usepackage{microtype}      
\usepackage{lipsum}		

\title{The Art of Social Bots: A Review and a Refined Taxonomy}

\author{
  \large Majd Latah\\
  \large Department of Computer Science\\
  \large Ozyegin University\\
  \large Istanbul, Turkey \\
  \large majd.latah@ozu.edu.tr \\
}

\begin{document}
\maketitle

\begin{abstract}
Social bots represent a new generation of bots that make use of online social networks (OSNs) as a command and control (C\&C) channel. Malicious social bots were responsible for launching large-scale spam campaigns, promoting low-cap stocks, manipulating user's digital influence and conducting political astroturf. This paper presents a detailed review on current social bots and proper techniques that can be used to fly under the radar of OSN defences to be undetectable for long periods of time. We also suggest a refined taxonomy of detection approaches from social network perspective, as well as commonly used datasets and their corresponding findings. Our study can help OSN administrators and researchers understand the destructive potential of malicious social bots and can improve the current defence strategies against them.
\end{abstract}


\section{Introduction}
Online social networks (OSNs) are popular platforms that connect users all over the globe. A botnet represents a group of agents (bots) that are managed and programmed to act in an organized manner. The term social botnet indicates a new generation botnet that utilizes OSNs as a command and control (C\&C) channel with minimal noise \cite{Burghouwt:2013}. A social botnet can be used for mis- and dis-information campaigns without being detected as pieces of software \cite{bruns2014metrics,bruns2014}. Social bots can also be used for benign purposes. For instance, on Wikipedia about 15\% of all edits are made by bots \cite{Steiner2014bots}. Even benign types of social bots can sometimes be harmful, for instance when they spread unverified information. This can be seen as a result to the lack of fast-checking in most automatic trading systems \cite{Ferrara:2016}. 

In 2010, \citeauthor{nappa2010take} designed a botnet using Skype protocol, which includes a fault-tolerant, encrypted communication. Moreover, it is firewall- and NAT-agnostic, which means that it does not require any additional network configuration. Koobface is another example of a malware that has proved successful in propagating through different social networks. The OSN propagation components of Koobface included several binaries, each of which has been developed to handle a particular OSN \cite{Baltazar:2009}. Nazbot \cite{Kartaltepe:2010}, named after Jose Nazario, is also an earlier social bot that uses Twitter as its C\&C channel. It uses an account named $upd4t3$ owned by the botmaster on Twitter to receive commands, which were encoded by base-64. Nazbot mainly employed Twitter's Really Simple Syndication (RSS) feed to receive botmaster's commands. However, Twitter does not provide RSS services anymore, and also tweets encoded with base-64 can be easily decoded \cite{he2017understand}.

In August of 2014, Twitter has reported that approximately 8.5\% of its users are bots, \cite{Subrahmanian:2016}. Moreover, \cite{zhang2011detect} showed that 16\% of these bots exhibit highly automated behavior. Recently, \cite{Cresci:2018} conducted a large-scale analysis of spam and bot activity in stock microblogs on Twitter. Their study showed that 71\% of retweeting users were classified as bots, 37\% of which were suspended by Twitter. 

\cite{Messias:2013} showed that Klout and Kred scores can be manipulated using simple automatic strategies. Data mining techniques can also be used to gain visibility and influence \cite{Bakshy:2011} \cite{Lee:2014} \cite{Suh:2010} \cite{Pramanik:2015}. \cite{zhang2016rise} showed that social bots can increase their the Kred and Retweet Rank influence scores of by purposely following each other. They also showed that Klout, Kred, and Retweet Rank increases as the number that a user is retweeted increases. In order to manipulate a targeted user’s influence score, social bots can either retweet a single tweet of the target user to make this tweet very popular or retweet many of target user tweets so that each tweet will be less popular. Bot-As-A-Service (BaaS) presents an emerging trend to provide a specific automated behavior on behalf of the user for a monthly price \cite{FM7090}. Such a service can be used to increase bot's influence score and to attract other legitimate users based on specific hashtags.


A 4-week competition was conducted by the Defense Advanced Research Projects Agency (DARPA) in February and March 2015, where the challenge was to identify influence bots on Twitter \cite{Subrahmanian:2016}. Competitors had to: (1) separate influence bots from other types of bots, (2) separate influence bots about topic \textit{t} from those about other topics and (3) separate influence bots about topic \textit{t} that sought to spread sentiments \textit{s} from those that were either neutral or that spread on opposite sentiment. Six teams (University of Southern California, Indiana University, Georgia Tech, Sentimetrix, IBM, and Boston Fusion), competed to discover 39 influence bots. The teams had no information about the number of bots included in the competition. DARPA botnet challenge suggested that a bot-detection system needs to be semi-supervised. That is, human judgments are required to improve detection of these bots. Visualization methods are also necessary to assets the experts to take better decisions about suspicious bots \cite{Subrahmanian:2016}.


In Table 1, we present a brief summary of existing real-world and research-based social bots. From this table, one can observe how the creators of social bot make use of additional  techniques to conduct successful malicious activities at both user-side and OSN-side. This may include further efforts at different stages of bot's life cycle. For instance, automatic CAPTCHA solver can be used to create fake  (sybil) accounts. Encryption techniques can be used in order to protect the confidentiality of the exchanged messages between the bot master and social bots from OSN's security defence. One can also observe that a great deal of recent research has focused on introducing advanced social bots through experiments or simulations, whereas few of them have been implemented in real network scenarios.



\begin{ThreePartTable}
\footnotesize

\begin{TableNotes}
\footnotesize
\item R: Real-world social bot, E: Experiment-based social bot, S: simulation-based social bot.
\end{TableNotes}

\begin{longtable}{| p{.15\textwidth} |  p{.13\textwidth} | p{.015\textwidth} | p{.12\textwidth} | p{.46\textwidth} |} 

\caption{A brief summary of existing social bots}
\label{tab:long} \\
\hline

\makecell{ Ref.} & \makecell{Name} & \makecell{R\\E\\S} & \makecell{Encoded\\Encrypted, or\\ Authenticated} & \makecell{Remarks} \\
\hline
\endfirsthead

\multicolumn{5}{c}%
{{\bfseries \tablename\ \thetable{} -- continued from previous page}} \\
\hline 
\makecell{ Ref.} & \makecell{Name} & \makecell{R\\E\\S} & \makecell{Encoded\\Encrypted, or\\ Authenticated} & \makecell{Remarks} \\
\hline
\endhead

\hline \multicolumn{5}{|r|}{{Continued on next page}} \\ \hline
\endfoot

\hline
\insertTableNotes
\endlastfoot

\cite{Baltazar:2009} & Koobface & R & \makecell{ Bitwise-ADD \\ \& bitwise-OR \\ operations } & \makecell{The propagation of zombies occurs as obfuscated URLs or\\by using re-directors.} 
\\
\hline

\cite{Kartaltepe:2010} & Nazbot & R & \makecell{base-64\\encoded text.} & \makecell{Uses RSS to receive encrypted tweets from the botmaster\\account on Twitter.}
\\
\hline

\cite{Xiang:2011andbot} & Andbot & S & \makecell{Base64 enc. \& \\public key enc.}  & \makecell{Uses URL Flux and avoids DNS sinkhole, malicious, commands\\ injection, IP blacklist and C\&C server shutdown attacks.}
\\
\hline

\cite{Lu:2011advancedp2p} & AHP2P Botnet 2.0 & S & \makecell{Message \\-Digest (MD5)} & \makecell{Consist of servant bots and client  bots, which attack\\target after they receive malware information from servant bot.}
\\
\hline 

\cite{Nagaraja2011stegobot} & Stegobot & S & \makecell{Image\\steganography} &  \makecell{It can only be detected using ad-hoc entropy measures
}
\\

\hline

\cite{Faghani2012socell} & Socellbot &  S & \makecell{No} & \makecell{Simulated on OSN graph and it infected smartphones.\\It is easily noticeable due to the generation of suspicious traffic.} \\
\hline

\cite{elishar2012organizational} & Organizational Socialbot & R & \makecell{No} & \makecell{Automatically sends friend requests in order to befriend\\with employees working in the targeted organization.\\They used Markov clustering and closeness centrality\\measure to cluster and to uncover leadership roles.} \\
\hline

\cite{Elyashar} & \makecell{Sophisticated\\ Organizational\\Socialbots} & R & \makecell{No} & \makecell{For each organization, they created a social-bot account\\that mimics legitimate users' behaviour and becomes friend\\with specific users in the targeted organizations. They were\\able to gain a successful infiltration with 50-70\%. 
} \\
\hline

\cite{Singh:2012} & Twitterbot & R & \makecell{No} & \makecell{It performs malicious attacks like browsing webpage,\\DoS attack, emailing files or gathering system information}   \\  \hline

\cite{Burghouwt:2013} & Twebot & R & \makecell{SSLv1} & \makecell{A variant of Nazbot, which polls a Twitter account\\every 20 seconds}
\\
\hline

\cite{Boshmaf:2013} & Yazanbot & R & \makecell{No} & \makecell{It includes an additional social bot-OSN channel and exploits\\triadic closure principle. It also uses HTTP-request\\templates to conduct  unsupported (illegal) API calls.}
\\
\hline

\cite{Cao:2013asp2p} & Asp2p & S & \makecell{Public key\\encryption \& \\authentication} &  \makecell{It includes enhancements over AHP2P 2.0 due to the key\\exchange communication schema.}
\\
\hline

\cite{zhang2013impact} & \makecell{spam distribution\\and influence\\ manipulation} & \makecell{S\\-\\R} & \makecell{No} & \makecell{They studied botnet-based spam distribution as an\\optimization problem. Also the influence manipulation\\was confirmed by real experiments.}
\\ 
\hline

\cite{yin:2014drsn} & Dr-snbot & R & \makecell{AES \& RSA} &  \makecell{The generation of network traffic makes it detectable\\with correlation and behavioural analysis. }
\\
\hline

\cite{Sebastian2014} & Graybot & E & \makecell{Diffie \\ Hellman} & \makecell{Attacks launched includes conversion of files\\ to shortcut, executables or keylogger. }
\\
\hline

\cite{paradise2014anti} & Targeted socialbots & S & \makecell{No} & \makecell{The most connected and page rank strategies were the most\\favorable for the defender. Whereas the most effective\\attack was randomly sprayed friend requests when the\\attacker knows the defense strategy.}
\\
\hline

\cite{Compagno:2015boten} & Boten ELISA  & E & \makecell{Unicode\\steganography,\\AES, RSA \\with SHA-2} & \makecell{ C\&C messages are piggybacked on the honest content of the\\victim posts on the OSN. It does not generate new\\traffic, which makes it unnoticeable by an infected user.}
\\
\hline

\cite{Pantic:2015covert} & Secret C\& C & S & \makecell{Encoding\\ map} & \makecell{Stegno system that allows secret communication\\ using only meta-data of the tweet.}
\\
\hline

\cite{Abokhodair:2015} & SyrianBot & R & \makecell{N.A} & \makecell{Used mis-direction and smoke screening to influence\\users' attention. It also included a peculiar hashtag,\\which is a combination of three  random letters.}
\\
\hline

\cite{He:2016} &  \makecell{WTAR-based \\ bots (Wbbot, \\Fbbot,Twbot) } & R & \makecell{ Data \\ Encryption \\ Standard \\ (DES)} &  \makecell{Used web test automation rootkit (WTAR) \\technique to carry out host and OSN activities. }

\\
\hline

\cite{Makkar:2017sociobot} & Sociobot & E & \makecell{Encoded} &  \makecell{Used Docker to simulate users. Each bot is a java\\ application that makes use of Twitter API. Epidemic\\models were used to validate and analyze the botnet.
}
\\
\hline

\cite{Echeverria:2017burst} & Bursty & R & \makecell{N.A} & \makecell{Included automatic CAPTCHA solver.\\It cannot be detected by supervised learning.}
\\
\hline

\cite{Echeverria:2017star} & Star Wars Bots & R & \makecell{N.A} & \makecell {Tweets only from Windows Phone sources and shows distinct\\textual features. Therefore, it can  be easily detected.}
\\
\hline

\cite{Wu:2018slbot} & SLBot & R & \makecell{Base64 en. \\public key \\ encryption} &  \makecell{Used steganographic techniques to hide command\\addresses in multimedia resources on OSNs.}
\\
\hline

\cite{Cresci:2018} & \makecell{Cashtag \\Piggybacking} & R & \makecell{N.A} & \makecell{Promotes low-value stocks by exploiting\\the popularity of high-value ones.}
\\
\hline

\end{longtable}
\end{ThreePartTable}

Previous research efforts were focused on coarse-grained taxonomy \cite{ramalingam2018fake,kumar2018false,kayes2017privacy,kaur2016survey,heydari2015detection}. However, in this work we provide more fine-grained taxonomy that leads to a better understanding of the state-of-the-art in social bot detection. In the next sections, we discuss different aspects to cover most essential aspects of the topic under consideration in order to see the big picture and to increase the awareness of OSNs' administrators, research community and the users who are susceptible to these attacks. This includes discussing different stealthy behaviours and techniques used by advanced social bots (Section 2). We also try to provide more detailed profiling of social bots in order to distinguish among different types at different levels (Section 3). Furthermore, we suggest a refined taxonomy of social-network based detection approaches (Section 4), as well as datasets used, and the corresponding findings in detail (Section 5). A brief summary of this paper is shown in Fig. 1.


\begin{figure*}[!b]
\centering
\includegraphics[width=170mm]{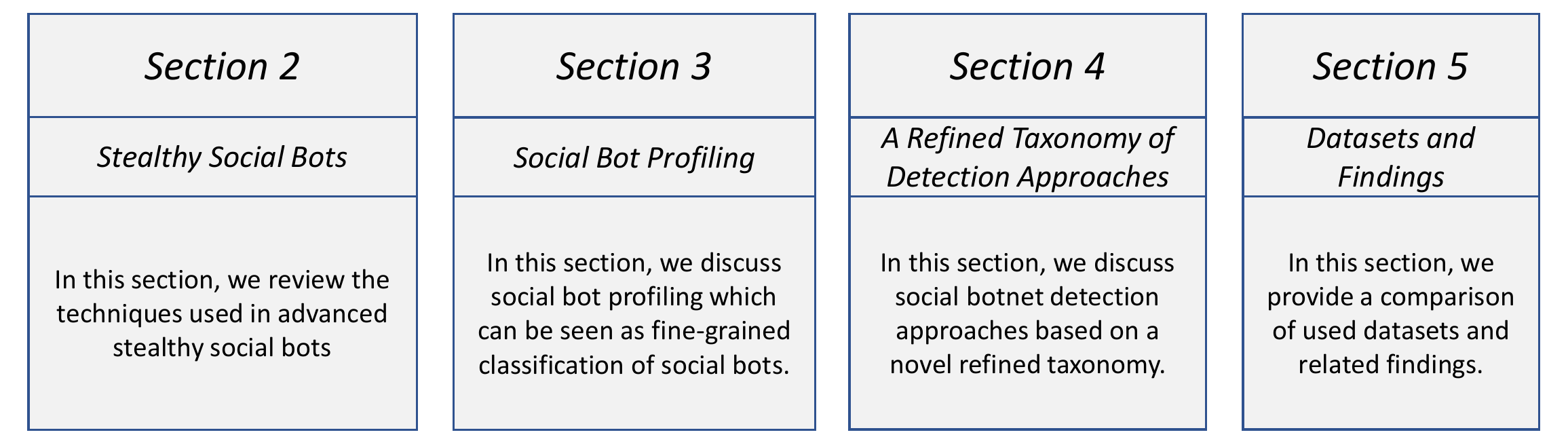}
\caption{A brief summary of our study.}
\end{figure*}

\section{Stealthy Social Bots}
In OSN settings, there is no communication between bots that do not exchange information with each other, therefore social bots do not add any connection end-points to communicate between themselves. Moreover, there is no additional communication beyond what that bot or other legitimate users already exchange, which results in a very stealthy design in compared with traditional bots \cite{Nagaraja2011stegobot}.

\cite{Cresci2017} highlighted the difficulties in distinguishing and taking an appropriate action against new generations of spam bots, which mimic the behavior of genuine users. Every bot had a profile filled in with detailed, yet fake, personal and other meta-data information. For instance, Global Positioning System (GPS) spoofing can be used to manipulate their geographical location. This type of bots were employed to support some candidates in the 2016 U.S. Presidential election \cite{FM7090}.

Furthermore, social bots can search the Web to find some useful information that may be used fill their profiles, and post these collected material at particular time periods, to imitate the users' temporal signature of producing and consuming information \cite{Ferrara:2016}. These bots were spotted during the last Mayoral election in Rome in 2014. A second group of social bots was used to promote a mobile phone application, called Talnts, to get in touch with professional talents in their location. A third group was used to advertise online products using spamming URLs. A human-to-spambot interaction was found in the first and third group. In other words, they found a set of legitimate users that interacts with these spam bots.

As the research community develops more advanced bot detection techniques, social bots also tend to be more sophisticated and find their way to evade detection. Therefore, from the botmaster’s perspective, new techniques are also being developed to evade the current OSN's defences using more stealthy, malicious yet cautious social bots that \textit{``act like human and think like a bot"} \cite{Ferrara:2016}. This fact shows that the developers of social bots follow the recent advances in defence techniques and figure out how to they can exploit them in their future bots. In the next subsections we discuss different techniques that have been used to develop very advanced stealthy bots.

\subsection{Malicious, yet cautious, use of OSN's API}
Bots constantly check OSN's API restrictions to guarantee that they would not be identified as bots and blocked. For instance Twitter's API restrictions include: (i) maximum number of requests per hour is 350; (ii) each user can follow up to 2,000 users; (iii) a user cannot follow more than 1,000 users per day. Twitter's Sample API, is a popular tool among researchers that returns a random sample of all public tweets generated on Twitter (1\%). Basically, Sample API selects tweets based on their timestamp. More precisely, any tweet whose ID is generated in the millisecond range of [657, 666] will be selected by the Sample API \cite{Kergl:2014}. This mechanism assumes that the millisecond-level information of the ID’s timestamp is random since humans don’t have any fine-grained control over the tweet’s time-stamp. However, social bots have the ability to achieve such fine-grained control that allows them to predict the time that Twitter receives their tweets and ultimately increase their chance to appear in the sample.

\cite{Morstatter:2016} confirmed the existence of strong linear relationship between the time the tweet is sent and the millisecond ID of the tweet ID. Accordingly, they designed an algorithm that adaptively measures the network lag between when the tweet is sent and when the tweet is received by Twitter, measured by the ID of the tweet. Moreover, Yazanbot \cite{Boshmaf:2013} used HTTP-request templates to conduct an unsupported API call by recording the exact HTTP requests that are used to carry out such a call. Moreover, a friend injection attack \cite{huber2010} allows a stealth infiltration of OSNs by tapping network traffic to extract authentication cookies and to submit requests on behalf of the legitimate user.

\subsection{Mimicking the behaviour of legitimate users}
\cite{GoncaHoca} introduced the concept of context-aware attacks, in which the attacker can mimic the behavior of legitimate user's traffic to be undetected for long periods of time. In OSN settings, a social bot can mimic a legitimate user's behavior, network structure or interest to appear as a legitimate user. For instance, \cite{Lee:2011} observed that social bots post on average only four tweets per day to appear as legitimate and ultimately evade detection by OSNs defences. Another interesting example from the Syrian social botnet showed that a bot-based opinion tweet, which mimics human interest, was retweeted by 237 human accounts \cite{Abokhodair:2015}. Moreover, social bots can search for the most influential people and send them spam messages based on their topic/event/keywords of interest to be more publicized. They can take advantage of recent advances of natural language algorithms to be effectively involved in popular discussions to gain legitimate users' attention \cite{Ferrara:2016}, and ultimately appear more legitimate than simple social bots.


\subsection{Inferring hidden information of legitimate users and organizations}
Social bots can infer information that is not explicitly mentioned or allowed to be accessed by other users in order to increase the likelihood of a successful infiltration \cite{Hwang:2012}. A social bot can be used for exposing private information (i.e. phone numbers and addresses) after a successful infiltration. A social bot can also interacts with a legitimate user's friends, which allows it to produce a coherent content with similar temporal patterns \cite{Ferrara:2016}. Moreover, a social bot may take the advantage of machine learning approaches to infer hidden attributes from public insensitive attributes in social networks. For instance, inferring age information based on available public user profiles \cite{zhang2016your} and interaction information \cite{Mei:2017}.




Furthermore, \cite{fire2016organization} introduced algorithms that can be used for constructing organizational crawlers, which collect data from Facebook network in order to gather public information on users who worked in a specific organization based on informal social relationships among employees of targeted organizations. By applying centrality analysis and machine learning techniques, they were able to restructure parts of the targeted organization and to discover hidden departments and leadership roles, among the many discoveries, in each organization.

Moreover, guided crawlers provide a proper balance between exploration and exploitation, which can be leveraged by social bots in order to target a certain organization \cite{bnaya2013bandit,bnaya2013social}. Such tools allow them to intelligently select organizational member profiles and monitor their activity to infringe the privacy of the targeted organization in general and the corresponding users in particular. As a result, the collected information can be used to falsely convince the targeted users that these crafted bots are more likely to be legitimate users rather than malicious accounts.

\subsection{Breaking automated-behavior detection techniques}
Koobface sends the CAPTCHA requests to other computers that are part of the botnet and waits for one of the humans at those other computers to enter the CAPTCHA information for it. If the account is locked down, Koobface can automatically verify the account and reactivate it. Koobface makes money by fraudulently driving traffic through a combination of mainstream affiliate advertising networks and criminal networks \cite{tanner2010koobface}.

Crowdsourcing can be used to break CAPTCHA as \cite{motoyama2010re} showed that companies charge \$1 per 1000 succesfully solved CAPTCHAs, and this process is automated using software APIs for uploading CAPTCHAs and receiving results. Overall, the service providers were reasonably accurate (86–89\%).

Ye et al. (2018) showed that they can break text-based CAPTCHAs using generative adversarial network (GAN). They evaluated their approach by applying it to 33 CAPTCHA schemes, including 11 schemes that are currently being used by 32 of the top-50 popular websites including Microsoft, Wikipedia, eBay and Google. This approach is highly efficient as it can solve a CAPTCHA within 0.05 second using a desktop GPU \cite{Ye:2018:Captcha}. 

Furthermore, \cite{bock2017uncaptcha} defeated Google's audio reCaptcha defence with over 85\% accuracy using multiple speech recognition tools. \cite{aiken2018poster} presented DeepCRACk, a deep learning model based on bidirectional recurrent neural network (BRNN) to automatically break audio CAPTCHAs with an accuracy of 98.8\%. The model was trained on 100,000 audio samples generated using SimpleCaptcha (an open-source CAPTCHA system) and showed the ability to break audio audio CAPTCHA in 1.25 seconds on average. \cite{sivakorn2016m}, on the other hand, investigated the role of Google’s tracking cookies, which are used in the CAPTCHA’s defenses. They have noticed that CAPTCHA system was aware of user's interacts with  Google services. They were able to fool the system after just 9 days of automated browsing across different Google services. \cite{sivakorn2016} used deep convolutional neural networks to break Google’s and Facebook's image-based CAPTCHAs. Their system achieved an accuracy of 70.78\% when applied to Google's reCaptcha challenges, while requiring only 19 seconds per challenge. It also showed an accuracy of 83.5\% when applied to Facebook's image CAPTCHAs. Overall, one can observe that text-, image- and even audio-based CAPTCHAs were all breakable using low-cost techniques, and therefore they can potentially be used by next generation social bots.

\subsection{Deceptive techniques for their malicious activity}
In practice, OSN's security defences may detect and therefore take down accounts associated with malicious activities. To solve this issue, social bots can find more deceptive techniques to maintain their malicious activities. For instance, Flashback a trojan targeting Mac OS X, instead of searching for Tweets from a specific user, it generates a hashtag, which is used to search for any tweet containing that particular hashtag. In this way, there is no single user account for Twitter to take down \cite{prince:2012Flashback} \cite{Lehtio:2015}. Spam bots also may include legitimate URLs in their posts to avoid  detection. This phenomenon is called \textit{legitimate URL inclusion attack} \cite{tan2013unik}. 

Moreover, \cite{cresci2019capability} developed a new model that produces variants of malicious bots, which need to satisfy a given criteria (i.e. evading detection) based on genetic algorithm approach. Online user behaviors were modeled (encoded) with DNA-inspired strings of characters representing the sequence of a user’s online actions. At each new generation a subset of the synthetic accounts is replaced with other synthetic accounts that are proven to be more similar to legitimate user accounts. The synthetic accounts generated were evaluated against a state- of-the-art detection technique and were able to evade detection. Apart from that, bots may choose to conduct their malicious activities during the night rather during the day to avoid reporting by other OSNs users.

\subsection{Deceptive social features}
Social bots can send a mixture of both malicious and legitimate messages. Hence in many cases they share the same sender and will naturally have exactly the same value of the sender’s social degree. This attack can be even more successful when the attacker uses a compromised account \cite{Gao2012towards}.
Social bots can also follow each other and retweet/answer others’ tweets. By maintaining a balanced following/followers ratio, individual bots can escape early detection in OSN platforms \cite{zhang2016rise}. Moreover, Yazanbot \cite{Boshmaf:2013} exploited triadic closure principle \cite{rapoport1953}, to increase the magnitude of the potential infiltration. They even showed that attackers can estimate the probability of a certain user to accept a friend request, given the total number of friends the user have and the number of mutual friends with the social bot \cite{boshmaf2011socialbot,Boshmaf:2013,paradise2014anti}. Therefore, in order to maximize the infiltration, sophisticated bots can choose a certain group of users with probability of accepting friend requests coming from social bots \cite{paradise2014anti}.

Furthermore, \cite{Hwang:2012} observed that the more friends a user has, the less selective he/she will be when filtering out social bot-based friendship requests. \cite{Wanger:2012} investigated the susceptibility of users who are under social bots attack. Their analysis interestingly showed that the attacks of social bots can be successful even when users are highly active and already have gained experience with social networks.

Apart from that, \cite{alvisi2013sok} empirically showed that well-known structural properties (i.e. popularity distribution among graph nodes, small world property and clustering coefficient) are not able to provide full-proof defence against sybil attacks. However, the conductance seems much interesting from OSN defence perspective as it takes much more effort to appear legitimate under strong attacks. The attacker enlarges every cut with some probability $p$, which increases the cut with minimum conductance, and ultimately the conductance of the  whole network.

\subsection{Sophisticated propagation of social bots}
As some OSNs such as Twitter, according its previous rules, suspends only the users that originate spam tweets without taking any action against those retweeting them. This idea motivated \cite{zhang2016rise} to represent spam distribution as a retweeting tree in which only the root originate spam tweets. Accordingly, all the social bots except the root can avoid detection by OSN defence. They also suggested a multi-objective optimization to minimize the delay to reach a maximum number of legitimate users at the minimum cost of spam distribution. 

\cite{faghani2018mobile} constructed a botnet propagation tree (BPT) from an OSN graph, after recruiting all the potential vulnerable nodes and removing the users that are not infected. The user can be infected as early as possible via the shortest path from the infiltrating node. They also studied various strategies that can be used to maximize the number of bots while minimizing the risk of being detected.


\subsection{Different methods to receive commands from the botmaster}
Malicious commands from the botmaster can be sent via unicast  or broadcast messages. In the former case,  the botmaster chooses the recipients of the commands from a list of bots. In the latter case, however, the recipients have to join one of different groups created by the botmaster, which sends only one message to several users in order to avoid or delay detection \cite{faghani2018mobile}. \cite{Nagaraja2011stegobot} used image steganography scheme over the Facebook social network to hide malicious information. Each bot gathers the information requested by the botmaster such as files matching search strings then encodes them as much as possible in a single image according to a detection threshold. In order to increase the robustness of their botnet, \cite{Wu:2018slbot} suggested that bots should receive the addresses of stored commands from an addressing channel, which can be any OSN that allows custom URLs and provides temporary storage where the address is Base64-encoded and signed by private key schema and hidden in a picture.


\section{Social Bot Profiling}
Many studies in the literature have focused merely on distinguishing legitimate users and malicious social bots \cite{Wang:2010a,Dickerson:2014,Dorri:2018,Ji:2016,Ferrara:2018,Besel:2018}. Other studies, however, suggested more fine-grained classification at different levels. For instance  \cite{Kartaltepe:2010}, in their client-side countermeasure, collected process-level information to distinguish benign and suspicious processes. 

In social graph-based defence approaches bots are often referred to as sybils (fake accounts) that are mainly used to forge other users' identities and launch malicious activities \cite{Ferrara:2016}. \cite{cao2013sybilfence} refereed to the sybils adjacent to the attack edges as entrance sybils and the other sybils as latent sybils.

Furthermore, social bots can also be categorized into different fine-grained classes. \cite{Chu:2012} defined a bot as an automated account used for spam purposes, whereas a cyborg represents a bot-assisted human or human-assisted bot. \cite{tavares:2013} categorized Twitter accounts into three groups: (1) personal, (2) managed and (3) bot-controlled accounts based on their tweet time intervals. Even though they may  involve in a malicious activity, all real users' accounts (such as human spammers, trolls, managed accounts) are not social bots. \cite{Stringhini:2010} also focused on different types of spam bots (Displayer, bragger, poster and whisperer). \cite{clark2016sifting} focused on three distinct classes of automated tweeting: robots, cyborgs and human spammers.  \cite{el2018supervised} studied verified users (cyborgs), human users, trends hijackers and promotional spam bots.

\cite{Lee:2011} suggested various categories of content polluters (Duplicate spammer, Duplicate @ spammer, malicious promoter and friends infiltrator). Duplicate spammers post nearly identical tweets with or without links. Duplicate @ Spammers they are similar to Duplicate spammers however they also abuse Twitter’s @username mechanism by randomly inserting a legitimate user’s @username. Malicious promoters include legitimate and more sophisticated tweeting about online business, marketing or finance. Friend infiltrators adopt \textit{``act legitimate and think malicious"} approach by abusing the reciprocity in following relationships.

\cite{Grier:2010} studied both carrier spammers and compromised accounts.
\cite{Abokhodair:2015} classified each bot into one of five categories (short-lived bots, long-lived bots, generator bots, core bots and peripheral bots) based on distinct behavioral patterns. \cite{song2015crowdtarget} analyzed retweets generated by three account groups: (1) normal, (2) crowdturfing, and (3) bots from black market sites. \cite{Benigni:2019} distinguished between promoted accounts and community influencers. \cite{fernquist2018} used the term bot to refer to accounts that show an automated behaviour, including automated social bots and sock puppets.

\cite{Echeverria:2018} focused on multiple bot types, which were obtained from a variety of social bot datasets. Their aggregated bot dataset contained over 1.5 million bots with all their available tweets. \cite{Yan:2013} investigated different types of social botnets namely standalone, appendix and crossover botnets. Standalone botnets are isolated from the normal online social network. Appendix botnets have only one direction following relationships. Crossover botnets have following relationships in both directions between bots and normal users. \cite{Beugenilmics:2018} distinguished between (organized- vs. organic-behavior), (pro-Trump vs. pro-Hillary vs none) and (political vs. nonpolitical). \cite{Gao2010} focused on detecting malicious (URL/posts). 

\cite{Gao2012towards} distinguished between spam and legitimate clusters. \cite{lee2011content} used graph mining to detect top largest campaigns (spam, promotion, template, news, and celebrity campaigns). Furthermore, honey-profiles can be used to capture behavioural patterns of social bots. For instance \cite{Stringhini:2010} in their honey-profile-based approach were able to identify two kinds of bot behaviors namely: stealthy and greedy bots. Greedy bots add a spam content in every message they send, whereas stealthy bots send messages that look legitimate, and inject them only once in a while \cite{Stringhini:2010}. On the other hand, \cite{Freitas:2015} suggested a reverse-engineering approach to understand different infiltration strategies of malicious social bots in Twitter.

Apart from these categories which mainly deal with malicious bot, \cite{oentaryo:2016} provided a new categorization that includes both benign and malicious bots. This behavior-based categorization included three main types namely broadcast, consumption and spam bots. A broadcast bot is managed by a particular organization or a group and mainly used for information dissemination purposes. Consumption bot also aims at aggregating contents from different sources and provides update services. Spam bot, on the other hand, is used for delivering malicious contents. 

As we have discussed above, social bot profiling is an essential step for many reasons: (1) detection approaches may focus on certain types of benign or malicious social bots while miss other types, (2) OSN administrators need to allow benign bots while mitigate the malicious ones and (3) other benign accounts such as cyborgs and managed accounts can be falsely detected as malicious bots, which yields undesirable high false-negative rates.

\section{Detection of Social Bots}
In this section, we focus on social network-based detection of malicious bots where the detection approaches fall into three main categories as shown in Fig. 2. Previous research efforts were focused on coarse grained taxonomy \cite{ramalingam2018fake,kumar2018false,kayes2017privacy,kaur2016survey,heydari2015detection}. However, this work provides a refined taxonomy that leads to a better understanding of the state-of-the-art in social bot detection. We also investigate the strength and weakness of these approaches, along with their effectiveness and weaknesses in real-world OSNs.

\begin{figure*}[ht!]
\centering
\includegraphics[width=175mm]{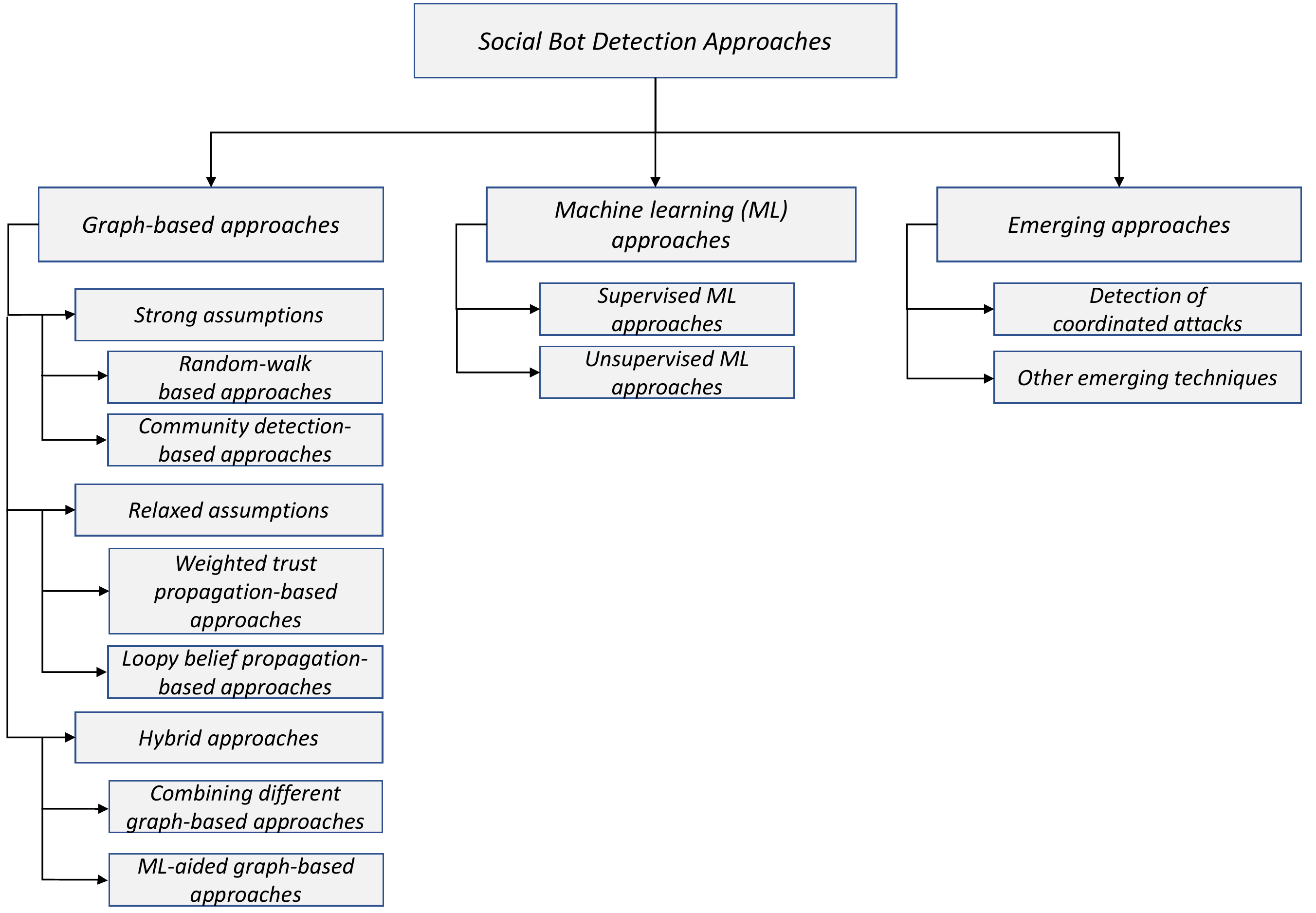}
\caption{A refined taxonomy of social bot detection approaches.}
\end{figure*}


\subsection{Graph-based approaches}


As we mentioned previously, in social graph-based defence approaches bots are often referred to as sybils (fake accounts), which are mainly used to forge other users' identities and launch malicious activities Ferrara et al. (2016). Graph-based approaches are based on three key assumptions. First, the cut between the sybil and honest regions is sparse. Second, the benign region is fast mixing, which implies that random walks of steps will end in a random edge in the honest region with high probability. Third, the social edges represent strong trust relationships, and therefore it is hard for the attacker to create links to the legitimate users (i.e. to the honest region) \cite{alvisi2013sok,SybilFuse}. Based on these three assumptions graph-based approaches can be categorized into the following groups: 

\subsubsection{Strong assumption-based approaches}
\paragraph{Random-walk-based approaches}
SybilGuard \cite{yu2006sybilguard} performs long random walks $\Theta$($\sqrt{n}$ log $n$) that should intersect with the suspect node in order to accept it as honest. SybilGuard makes use of the following observations, which hold only when the number of attack edges is limited: (i) the average honest node’s random route is highly likely to stay within the honest region and (ii) two random routes from honest nodes are highly likely to intersect within the random walk. 
 
SybilLimit \cite{yu2008sybillimit}, on the other hand, uses multiple instances ($\sqrt{m}$) of short random walks O(log $n$) to sample nodes from the honest set and accepts a node as honest when there is an intersection among the last edge of the random routes of the verifier and the suspect, and the intersection satisfies a balance constraint. It is obvious that SybilGuard performs intersections on nodes, whereas SybilLimit applies intersections on edges. Both SybilGuard and SybilLimit are high vulnerable when high degree nodes are compromised. Therefore, one can conclude that these two approaches are more effective in defending against malicious users than defending against compromised honest users. They also include a pre-processing step in which nodes with degrees smaller than a pre-defined threshold were removed from theses approaches. As a result, a large number of nodes will be pruned \cite{mohaisen2010measuring} due to the fact that social networks often have a long-tail degree distribution \cite{clauset2009power,gong2012evolution}. Moreover, it can result in high false positive/negative rates according to how the OSN operator treats these pruned nodes \cite{gao2015sybilframe}.


Fast mixing with the honest region implies the absence of small cuts whereas the slow mixing between honest and dishonest implies that one can compute the bottleneck cut between honest and sybil nodes to infer the labels of each node \cite{danezis2009sybilinfer}. SybilInfer considers the disturbance in fast-mixing between the honest and dishonest region as a faint bias in the last node of a short random walk. Unlike SybilGuard and SybilLimit, SybilInfer depends primarily on the number of colluding malicious nodes, not on the number of attack edges and it also performs independently of the topology of the sybil region \cite{danezis2009sybilinfer}. 
 
SybilResist \cite{ma2014sybil} consists of four main parts and four stages: (1) performing random walks O(log $n$)  starting from a known node. Then, selects the nodes with a frequency higher than a pre-defined threshold. These nodes are treated as honest nodes (2) sybil identification to identify the suspect nodes based on comparing with the mean and standard deviation of the previous random walks; (3) walk length estimation to set the initial value for Sybil region detection algorithms and (4) detecting the Sybil region among the detected nodes. 
The main drawback here is that every time the structure of the sybil region changes, the performance changes according to the length of random walks used in each step. Furthermore, SybilResist is also high vulnerable when high degree nodes are compromised.

SybilRank \cite{cao2012aiding}, uses short random walks to distribute initial scores from a set of trusted benign seeds. It consists of three steps: 1) trust propagation, 2) trust normalization, and 3) ranking. In the first stage, SybilRank propagates trust from the trust seeds via O(log $|V|$) power iterations. The second stage includes normalization of the total trust of every node with its social degree. In the last stage, SybilRank generates a ranked list based on the degree-normalized trust with non-sybil users on top.
SybilRank exploits the mixing time difference between the non-Sybil region and the entire graph to distinguish sybil from non-sybils. However, it requires that the sybil region have a longer mixing time than the non-sybil region where the early-terminated power iteration result in a gap between the degree-normalized trust of non-sybils and sybils. SybilRank makes use of the paradigm of innocent by association, in which an account interacting with a legitimate user is considered itself legitimate. This approach relies on the assumption that legitimate users refuse to interact with unknown accounts, which was proven to be unrealistic by recent experiments \cite{Stringhini:2010,boshmaf2013design,Elyashar}, and therefore allows undesirable high false-negative rates \cite{Ferrara:2016}.

Previous approaches such as SybilGuard, SybilResist, SybilLimit, and SybilInfer are not scalable to large-scale OSNs and moreover they do not utilize information about known Sybil labels. In fact, these approaches only incorporate one labeled benign node, which makes them not resilient to label noise and can only allow one node to be identified at a time and therefore finding the Sybil region will be time-consuming due to the fact they need to examine the whole graph \cite{gong2014sybilbelief,ma2014sybil}. Moreover, \cite{liu2015exploiting} showed that these systems are vulnerable to temporal attacks, which can result in dramatic gains over time and significantly degrade the security guarantees of the system in distributed sybil defenses such as SybilLimit or can escape detection against centrally controlled sybil defenses such as SybilInfer. To attack random-walk based approaches, the intruder has to choose (a) a good attack strategy and (b) gain some additional information which includes: (1) topology, (2) trusted nodes, and (3) untrusted nodes to efficiently guide the creation of attacking edges, which allows him to obtain enough trust \cite{honer2017minimizing}. Additionally, a trust leak in the first iteration can be exploited by a greedy strategy that iteratively adds those attacking edges with the largest gain/leak per attacking edge in order to deceive the detector by increasing the trust values of sybil nodes \cite{honer2017minimizing}.






\paragraph{Community detection approaches}
assume that OSNs under a social bot attack can be seen as two tightly-connected communities, namely sybil and non-sybil communities.


First, we need to show how we can capture the community structure for a given social graph. Interestingly, the community structure can be inferred by maximizing the modularity \cite{newman2004finding}.
\cite{clauset2004finding,gao2015sybilframe,SybilFuse,viswanath2011analysis} used modularity to verify (or falsify) whether a large-scale Twitter network can be seen as two communities. \cite{clauset2004finding} experiments on real networks showed a significant community structure. However, \cite{gao2015sybilframe,SybilFuse} falsified this observation as they found that the partition consisting of the benign and sybil regions only has a very low modularity. This assumption fails due to the fact that a considerable portion of the sybil nodes are $isolated$, and that the number of attack edges per sybil node is high \cite{gao2015sybilframe,SybilFuse}.

Conductance is also a well-known metric to find good communities. \cite{leskovec2008s} informally defined the conductance of a set of nodes as \textit{``The ratio of the number of cut edges between that set and its complement divided by the number of internal edges inside that set"}, where lower conductance indicates stronger communities.
Moreover, the conductance seems an interesting metric from OSN defence perspective as it takes much more effort to appear legitimate under strong attacks. As the attacker enlarges every cut with some probability $p$, which increases the cut with minimum conductance and, ultimately the conductance of the  whole network \cite{alvisi2013sok}.

\cite{mislove2010you} proposed a community detection approach based on a greedy approach to maximize the normalized conductance. This approach was applied in the following studies to detect sybil communities. In \citeauthor{viswanath2011analysis}'s work, they selected Mislove’s algorithm, however with the different conductance metric, which is the mutual information between pairs of rankings at all possible cutoff points (i.e. the boundary between the partitions). Ranking nodes can be achieved based on how well the nodes are connected to a trusted node (i.e. based on SybilGuard or SybilLimit). This approach outperformed both SybilGuard and SybilLimit. This is due to the fact that the number of Sybil nodes added is lower than the bound provided by these two approaches.

\cite{tan2013unik} proposed sybil defense-based spam detection scheme (SD2) based on user graph, which is formed by combining the social graph and the user-link graph. In this scheme, community detection based on \citeauthor{viswanath2011analysis}'s work is applied to the user graph to rank nodes where non-sybil nodes will have higher ranks and sybil nodes will have lower ranks. Finally, a cutoff is applied to ranked nodes based on the conductance ratio to identify sybil nodes as spammers. 


\cite{cai2012latent} suggested that it is not enough to partition the network into tightly-connected communities. However, these communities must $simultaneously$ be studied to figure out how they are connected with the rest of the OSN. Therefore, they proposed a latent community model for partitioning a graph into sub-networks. They applied Bayesian inference approach for learning the LC model, as well as associated Markov chain Monte Carlo (MCMC) algorithms. As the communities are learned, they are simultaneously positioned in a latent, Euclidean space so that communities with tight interconnections are positioned more closely than communities that are loosely connected. Therefore, attack communities tend to be seen as outliers based on the proper latent positioning. LC approach outperformed \cite{danezis2009sybilinfer}; however, it does not work well under a tree-topology attack (a tree has very low density).




%

Community detection methods need to take into account that a smart attacker may mimic the community structure of some legitimate users to appear as a human account \cite{Ferrara:2016}. Additionally, the attacker may set-up the attack links to deceptively appear as a part of a trusted node’s local community and again appear more legitimate \cite{viswanath2011analysis}.

Moreover, \cite{viswanath2011analysis} found that the effectiveness of community detection approaches depends on the level of community structure present in the non-sybil region of the OSN. For instance, the performance increases in OSNs that have low community structure and decreases in those with high community structure. Accordingly, OSNs appear to have many local communities or groups, rather than two tightly-connected communities \cite{cai2012latent}.

%

\subsubsection{Relaxed assumption-based approaches}

First, the previous approaches bound the accepted sybils to the number of attack edges based on social graph properties and structure. As the third assumption suggests that the number of attack edges is limited.
However, \cite{sridharan2012twitter,boshmaf2011socialbot} found that sybils are able to create a larger number of attack edges. In other words, the limited-attack-edges assumption does not hold in real-world OSNs. \cite{yang2014uncovering} showed that RenRen, the largest social networking platform in China, does not follow this assumption, which also implies that real-world social networks may not necessarily represent strong trust networks. 

Second, \cite{Ghosh:2012} examined the link farming phenomenon in the Twitter and they showed that specific users, called social capitalists, have a tendency to follow back  any account who follows them, in order to promote their content and to increase their social capital. Spammers, on the other hand, can exploit this behavior to farm links and promote their malicious content. This phenomenon can be seen as another evidence of that real-world social networks may not necessarily represent strong trust networks and therefore honest region may not be easily separable from the sybil region \cite{gao2015sybilframe}. 

Third, \cite{mohaisen2010measuring} empirically observed that the mixing time of real-world social graphs is longer than the theoretical anticipated value. The fast mixing assumption postulates the existence a small cut, a set of edges that together have small stationary probability and whose removal disconnects the graph into two large sub-graphs \cite{danezis2009sybilinfer,gao2015sybilframe}. More precisely, an OSN with weak trust does not show the fast mixing property \cite{mulamba2016ybilradar} and we already have shown two cases in which real-world social networks do not necessarily represent strong trust relationship. Therefore, in particular, we need better approaches that relax these assumptions in order to detect the sybil accounts that already have obtained social edges to real users (i.e. to the honest region). 

\paragraph{Weighted trust propagation-based approaches}
The following approaches propagate trust from a weighted social graph via power iterations (i.e. a PageRank like approach). SybilFence \cite{cao2013sybilfence} make use of the observation that most of fake users receive a large number of legitimate users' negative feedback, such as the rejections to their friend requests. SybilFence reduces the weights of social edges on users that have received negative feedback, which, in turn, limits the impact of sybils’ social edges. SybilFence adapts SybilRank approach to its weighted defense graph. It mainly consists of three steps: (1) trust propagation, (2) trust normalization, and (3) ranking users based on their degree-normalized trust. SybilFence is resilience to Sybils’ flooding requests and it outperforms SybilRank due to taking advantage from the consideration of negative feedback. However, SybilFence assumes that the non-sybil region is well connected after social edges discount, which can be unrealistic assumption in real-world social networks such as RenRen   \cite{yang2014uncovering}.

VoteTrust \cite{xue2013votetrust} is a global voting-based approach that combines link structure and users' feedback to detect sybils in OSNs. It assumes that the attacker can send many friendship requests to the honest users, however it can receive a large amount of rejections \cite{xue2013votetrust,koll2017thank}. The invitation graph is separated into two graphs: (1) the link initiation graph, which models the link initiation interactions as a directed graph and (2) the link acceptance graph, which models the link acceptance interactions as a weighted-directed graph. The weight value represents whether a user $u$ accepts or rejects the request. VoteTrust introduces two key techniques against collusive rating: (1) trust-based votes assignment and (2) global vote aggregating. The first step includes vote propagation from trusted seeds, which assigns low voting capacity for sybil nodes and thus prevents them from being involved in collusive rating. Then, it evaluates a global rating for each node by aggregating the votes all over the OSN. VoteTurst essentially ignores the votes from nodes of very low capacity if their trusting rating falls below a threshold, which forms a trade-off between collecting most votes and sybil community size. Interestingly, as sybil community grows, the vote capacity decreases due to sharing a fixed number of incoming links. VoteTrust can detect attackers even when sybil nodes are isolated and have many attack edges. However, VoteTrust can be invaded by tricking a few honest nodes into sending requests to sybil nodes or by sending requests to friends of already established victims \cite{koll2017thank}.

\paragraph{Loopy belief propagation-based approaches}
SybilBelief \cite{gong2014sybilbelief} utilizes information from a \textit{small set} of known benign and/or sybils users. It does not use random walks, instead it forms the problem of finding sybil accounts as a semi-supervised learning problem, in which the goal is to propagate reputations from a small set of known benign and/or sybils to other users along the social connections between them. SybilBelief mainly relies on the Markov Random Fields and Loopy Belief Propagation, and performs orders of magnitude better than SybilLimit and SybilInfer in terms of both the number of accepted sybil nodes and the number of rejected benign nodes. However, the number of accepted Sybil nodes increases dramatically when the labeled benign and Sybil nodes are highly imbalanced. SybilBelief adopts Loopy Belief Propagation (LBP) to make inferences about the posterior information. However, for social networks with loops LBP approximates the posterior probability distribution without theoretical convergence guarantees. Another limitation is that LBP-based methods are sensitive to the number of iterations that the methods run \cite{wang2017sybilscar}.

\paragraph{Hybrid approaches}

\subparagraph{Combining different graph-based approaches} 


SybilRadar \cite{mulamba2016ybilradar} is a weighted trust propagation-based approach to protect OSNs with weak trust relationships. It consists of three stages. The first stage includes computing of similarity values between a given pair of nodes based on Adamic-Adar metric. The second step represents a refinement of the previous step using another similarity metric, which is the Within-Inter-Community metric (WIC). The Louvain method \cite{blondel2008fast} is utilized to detect the corresponding communities, which are fed to the WIC similarity metric computation. The resulting similarity values are assigned to the social graph edges as their weights, which are used to ensure that a big fraction of the total trust will be distributed to legitimate nodes rather than to Sybil nodes. In the third stage, trust values are obtained using a modified short random walk O(log $n$) on the weighted social graph. SybilRadar showed a good performance even when the number of attack edges increases in compared to SybilRank. On the other hand, the increase in the size of the Sybil region affects the performance of both SybilRadar and SybilRank. However, SybilRadar is less sensitive to the size of the Sybil region.

Graph-based approaches can be seen as iteratively applying a local rule to every node in a weighted social network. \cite{wang2017sybilscar} proposed SybilSCAR a local rule that updates the posterior knowledge of a node by combining the influences from its neighbours with its prior knowledge. Their proposed approach is able to tolerate label noise and guarantees the convergence on real-world OSNs by a making use of a linear approximation of the multiplicative local rule and avoiding maintaining neighbour influences associated with every edge. They associate a weight with each edge, which represents the probability that the two corresponding users have the same label based on the homophily strength. Suppose a node $v$ is node $u$'s neighbor, then the local rule models $v$’s influence to a node $u$’s label as the probability that $u$ is a sybil, given $v$’s information alone. SybilSCAR achieves better detection accuracy than SybilRank and SybilBelief. SybilSCAR significantly outperformed SybilRank in terms of accuracy and robustness to label noise, and SybilBelief in terms of scalability and convergence.

\cite{zhang2018sybil} proposed SybilSAN, a two-layer hyper-graph model to fully utilize users’ friendships and their corresponding activities in an OSN in order to enhance the robustness of the sybil detection in the presence of a large number of attack edges. They also introduced a new sybil attack model in which sybils can launch both friendship and activity attacks. They used Markov chain mixing time to derive the number of rounds needed to guarantee that the iterative algorithm terminates. They decompose the graph into three sub-graphs: (1) the friendship graph; (2) the activity-following graph; (3) the user-activity graph. For each sub-graph, they designed a random walk to propagate trust independently. Finally, they present a unified algorithm to couple these three random walks to capture a mutual reinforcement relationship between users and activities. Under different attack scenarios SybilSAN outperformed SybilWalk, SybilSCAR and SybilRank. However, its performance degrades when the sybil region is split into a larger number of disconnected clusters.




SybilBlind \cite{SybilBlind} is a hybrid detection framework that does not rely on a manually labeled data. The main principle is to randomly sample users from the social network, and using them as a training set. SybilBlind mainly consists of three stages: (1) randomly sampling a noisy training set, (2) the noisy training set is then used as an input to state-of-the-art sybil detection method (SybilSCAR), and (3) finally previously obtained results are aggregated based on homophily-entropy aggregator (HEA). SybilBlind outperformed community detection \cite{blondel2008fast}, SybilRank, SybilSCAR (with sampling as training data) and showed comparable performance with SybilBelief.

\subparagraph{ML-aided graph-based approaches} 

SybilFrame \cite{gao2015sybilframe} is a multi-stage classification approach that makes use of the attributes of an individual node and the correlation between connected nodes. It consists of two stages of inference. The first stage includes exploring the dataset and extracting useful information, in order to compute node prior information and edge prior information based on SVM approach and similarity metrics respectively. This prior information along with a small set of nodes whose labels are known will be fed into next stage, which represents the posterior inference layer where the correlation between nodes is modeled using pairwise Markov Random Field.

Integro \cite{boshmaf2015integro} extracts content-based features and use them to train a classifier that predicts potential victims of sybil attacks. Then, the edges in the social graph are given weights according to their adjacency to the potential victims. The ranking is then accomplished via a modified random walk. Both SybilFrame and Integro make use of structure- and content-based information to achieve better detection accuracy.


SybilFuse \cite{SybilFuse} utilizes a collective classification approach and includes the following two steps: (1) training local classifiers to compute local trust scores for nodes and edges, (2) propagating the local scores via weighted random walk or weighted loopy belief propagation mechanisms. The experimental results showed that weighted loopy belief propagation performs better than weighted random walks in all settings and it also achieved the best performance among all other evaluated approaches which includes: SybilRank, SybilBelief, SybilSCAR, Integro in ranking both isolated sybils and sybils in the largest connected component. Table 2 provides a comparison of graph-based detection approaches.


\footnotesize
\begin{longtable}{| p{.16\textwidth} | p{.09\textwidth} | p{.34\textwidth} | p{.32\textwidth} |} 

\caption{Comparison of graph-based social bot detection approaches}

\label{tab:long} \\
\hline

\makecell{Ref.} & \makecell{Model} & \makecell{Features} & \makecell{Other Details}  \\
\hline
\endfirsthead

\multicolumn{4}{c}%
{{\bfseries \tablename\ \thetable{} -- continued from previous page}} \\
\hline 
\makecell{Ref.} & \makecell{Model} & \makecell{Features} & \makecell{Other Details} \\
\hline
\endhead

\hline \multicolumn{4}{|r|}{{Continued on next page}} \\ \hline
\endfoot

\hline
\endlastfoot

\cite{yu2006sybilguard} & \makecell{SybilGuard} & \makecell{Relies on intersections between of verifiable\\random walks. Suffers from high false\\negatives. Requires knowledge on the \\complete network topology. It depends on\\the number of attack edges.} & \makecell{Accepts O($\sqrt{n}$ log $n$) sybils per attack edge.\\ The length of the random walk is\\ O($\sqrt{n}$ log $n$). It contains O($\sqrt{n}$)\\independent samples drawn roughly\\from the stationary distribution.} \\

\hline

\cite{yu2008sybillimit} & \makecell{SybilLimit} & \makecell{Provides very weak guarantees when\\high degree nodes are compromised\\and can protect the system only\\for f$<$0:01 compromised nodes.} & \makecell{SybilLimit accepts O(log $n$) sybils per attack\\ edge. Uses multiple instances ($\sqrt{m}$) of\\short random walks O(log $n$) to sample\\nodes from the honest set, where $m$ \\denotes the number of edges amongst\\the honest nodes.
} \\

\hline

\cite{danezis2009sybilinfer} & \makecell{SybilInfer} & \makecell{Uses Bayesian inference to detect approximate\\cuts between honest and sybil node regions.\\Compromised nodes get no advantage by\\connecting any additional sybil nodes.\\Depends primarily on the number of colluding\\malicious nodes not on the number of attack\\edges. Also independent of the topology\\of the adversary region. } & \makecell{The length of the random walk is O(log $\left|V\right|$).\\It does not specify
any upper-bound\\guarantee on false rates and
provides\\weaker guarantees in practice since\\ the adversary can inject sybils into \\a region undetected as long as the\\threshold $E_{xx}$ is not exceeded.} \\
\hline

\cite{viswanath2011analysis} & \makecell{Community\\ detection} & \makecell{Community detection based on Mislove's \\algorithm and the mutual information\\between pairs of rankings at all possible\\cutoff points as a conductance metric.} & \makecell{A smart attacker may mimic the community\\structure of other legitimate users or set-up\\the attack links to deceptively appear as\\part of a trusted node’s local community} \\

\hline

\cite{cai2012latent} & \makecell{Latent\\community\\ model} & \makecell{The communities are learned and positioned in\\a latent Euclidean space so that communities\\with tight interconnections are positioned\\more closely than communities that\\are loosely connected } & \makecell{LC approach does
not work well\\under a tree-topology attack\\(a tree has very low density)} \\

\hline




\cite{cao2012aiding} & \makecell{SybilRank} & \makecell{Constructs a defense graph with\\reduced weights on attack edges.} & \makecell{Requires that sybil region
have a longer\\mixing time
than the non-sybil region\\
and assumes that legitimate
users refuse\\to interact with
unknown accounts.} \\

\hline


\cite{xue2013votetrust} & \makecell{VoteTrust} & \makecell{Combines link structure and users' feedback\\to detect sybils and includes (1) trust-based\\votes assignment and (2) vote aggregating} & \makecell{As sybil community grows, the vote capacity\\decreases due to sharing of a fixed number of \\incoming links. It can be invaded by tricking\\ a few honest nodes into sending requests\\to sybils or by sending requests to friends\\of already established victims} \\
\hline


\cite{gong2014sybilbelief} & \makecell{SybilBelief} & \makecell{
Relies on the Markov Random Fields\\and Loopy Belief Propagation} & \makecell{ The number of accepted sybils increases\\ dramatically when the labeled benign\\and sybil nodes are highly imbalanced} \\
\hline

\cite{ma2014sybil} & \makecell{SybilResist} & \makecell{Detects the sybil region, a sub-graph of sybil\\nodes which, doesn’t have a small cut.} & \makecell{Random walks of O(log $n$). However\\its performance changes according to the\\length of random walks used in each step.} \\

\hline

\cite{boshmaf2015integro} & \makecell{Integro} & \makecell{Weighted graph-based on the adjacency\\to potential victims} & \makecell{Makes use of structure- and content-based\\features to achieve better detection} \\

\hline

\cite{gao2015sybilframe} & \makecell{SybilFrame} & \makecell{Computes node prior information and edge\\prior information based on SVM approach and\\similarity metrics respectively. Then, the \\correlation between nodes is modeled using \\pairwise Markov Random Field.} & \makecell{Its performance depends on the \\accuracy of external classifier} \\

\hline

\cite{mulamba2016ybilradar} & \makecell{SybilRadar} & \makecell{A hybrid approach in which weights of the\\edges are assigned according to similarity\\values based on inter-community similarity\\metric. Thereafter, trust values are\\obtained using a modified short random walk.} & \makecell{The increase in the size of the sybil\\region affects its performance. However,\\SybilRadar is less sensitive to the size\\of the Sybil region than SybilRank.
} \\

\hline



\cite{wang2017sybilscar} & \makecell{SybilSCAR} & \makecell{Unify random walk-based methods and loop\\belief propagation-based methods and updates\\the posterior of a node by combining its\\neighbours influence with its prior knowledge} & \makecell{Each edge has a weight that represents\\the probability that the two corresponding\\nodes have the same label based on the\\homophily strength} \\

\hline

\cite{SybilBlind} & \makecell{SybilBlind} & \makecell{Randomly sampling a noisy data, which is\\used as an input to a sybil detection system\\(SybilSCAR) to detect sybils, and then\\ aggregates the results from multiples sampling\\ trials based on homophily-entropy aggregator} & \makecell{The main advantage is that it \\does not rely on a manually\\labeled
training set} \\

\hline

\cite{SybilFuse} & \makecell{SybilFuse} & \makecell{Propagates the local trust scores via\\weighted random walk or weighted\\
loopy belief propagation mechanisms} & \makecell{Local classifiers to compute local\\trust scores for nodes and edge.\\
However, weighted loopy belief\\
propagation performs better. } \\

\hline

\cite{zhang2018sybil} & \makecell{SybilSAN} & \makecell{Three random walks on three sub-graphs\\to capture a mutual relationship \\between users \& activities} & \makecell{Shows a low performance when the sybil\\region is split into a large number of \\disconnected clusters.} \\

\hline
\end{longtable}

\normalsize

\subsection{Machine learning approaches}
Machine learning-based social bot detection approaches can be categorized into the following two groups:

\subsubsection{Supervised machine learning approaches}

Supervised Machine Learning (ML) approaches focus on various features that allow distinguishing between human and bot accounts. User meta-data and content features have been proven to be the most predictive and the most interpretable ones that can be compared with that of legitimate users to infer whether an account is likely a social bot or not \cite{Fazil:2018,Onur:2017,fernquist2018}.


\cite{Wang:2010a} applied supervised ML approach based on three content-based and three graph-based features to distinguish between human and spam bots accounts. Well-known classification approaches, such as decision tree (DT), neural network (NN), support vector machines (SVM), and Naive Bayesian (NB) were applied to identify spam bots on Twitter. NB showed the best results as it is more robust against noisy data.

\cite{Stringhini:2010} employed classification (Random Forest- RF) approach based on the following features: friend requests ratio, URL ratio, message similarity, friend choice, messages sent and friends number. They also aggregated different spam campaigns according to URLs that advertised the same page.


\cite{Lee:2011} utilized Expectation Maximization (EM) approach to group malicious bots into four categories (Duplicate spammer, @ spammer, malicious promoter and friends infiltrator) based on Expectation Maximization (EM) approach. Then employed naive bayes, logistic regression, support vector machine (SVM), and ensemble learning based on RF to predict whether a candidate Twitter account is a social bot or not. They utilized features that belong to the following four groups: 1) user demographics, 2) user friendship networks, 3) user content and 4) user history. The experimental results showed that boosting of RF achieved the best results.

\cite{Chu:2012} proposed a four-stage approach to distinguish between human, bot and cyborg accounts. Their system consists of four stages: 1) computing corrected conditional entropy to detect periodic or regular timing, 2) spam detection based on text-based Bayesian classification, 3) computing the bot deviation from the normal human distribution based on account-related features, and 4) finally decision making based on random forest approach.

\cite{yang2013empirical} proposed 10 novel features including (3) graph-based, (3) neighbor-based, (3) automation-based and (1) timing-based feature to infer whether a Twitter account is genuine or spambot. Graph-based and neighbor-based features were useful to find malicious bots who attempt to evade profile-based features by adjusting their own social behaviors. Whereas automation-based and timing-based features were proposed to detect social bots that attempt to evade content-based features by increasing the number of their human-like tweets. These novel features were evaluated using four different machine learning classifiers namely Random Forest (RF), Decision Tree (DT), Bayes Net (BN) and Decorate (DE). Accordingly, one can observe that RF was widely employed in supervised ML-based approaches and achieved the best results in \cite{Stringhini:2010,Lee:2011,Chu:2012,yang2013empirical}.




\cite{tavares:2013} applied Naive Bayes classifier to distinguish between personal, managed and bot based on distribution of tweet time interval. They found that the power-law distributions of inter-tweet delays shows a clear difference across these three mentioned classes. They also showed that it is possible to predict the probability distribution of a user’s the time delay with a  coefficient of determination of $\sim$ 0.7.

\cite{Dickerson:2014} employed ensemble of classifiers (including support vector machines (SVM), Gaussian naive Bayes, AdaBoost, gradient boosting, random forests (RF), and extremely randomized trees) based on tweet syntax, tweet semantics (at the individual user or neighborhood level), user behavior and network-centric user features. 
They also applied sentiment analysis on a per-user basis over a variety of topics based on \citeauthor{benamara2007sentiment,subrahmanian2008ava,barbosa2010robust,agarwal2011sentiment}. Moreover, they identified topics discussed by various Twitter users by employing latent Dirichlet allocation (LDA) to classify individuals as either bots or humans. They also employed kernel principal component analysis (PCA) for de-noising and dimensionality reduction where AdaBoost performed best on the reduced feature set, and gradient boosting performed best on the full feature set. Their sentiment features improved the accuracy of the classifier. In particular, the Area under the ROC Curve (AUC) increased from 0.65 to 0.73. Interestingly, they found that when a user’s fraction of tweets with sentiment between 0.5 and 0.9, he/she is much more likely to be a human than a bot.

\cite{oentaryo:2016} utilized four classifiers, namely NB, RF, SVM and logistic regression (LR) to distinguish between human accounts and three types of bots namely broadcast, consumption and spam bots. They considered profile-, follow-, static (i.e., time-independent) and dynamic tweet-based (i.e., time-dependent) features. Overall, LR and SVM showed the best results.


\cite{Fazil:2018} identified six newly features and two redefined features. The six new features include one content-based, three interaction-based, and two community-based features. The redefined features are content-based. These features were fed to RF, DT and Bayesian network classifiers to distinguish between automated spammers and legitimate users. They found that interaction- and community-based features are the most effective features for spam detection, whereas metadata-based features are the least effective ones. The interaction-based features, focused on the followers of a user, rather than on the ones he/she is following because these features cannot be determined by the user. The experimental results showed that RF achieved the best results over DT and Bayesian network. This approach can be seen as a hybrid approach as it depend on graph-based features as well as content-based features.


\cite{Davis:2016,Onur:2017} proposed BotOrNot, in which they applied RF approach based on more than 1,000 features using user profile-, friending- and network-, temporal-, content- and sentiment-based features extracted from interaction patterns and content. BotOrNot mainly computes the $botness$ of a certain user (i.e. the likelihood that an account is a bot). Various off-the-shelf classification methods were tested, including logistic regression, DT, RF and AdaBoost. However, the experimental results showed that RF is the most accurate to produce bot-likelihood scores. \cite{grimme2018changing}, on the other hand, showed that the average user's score of BotOrNot drops significantly when bots start spreading tweets through the network. This is due to the fact that BotOrNot’s features include the number of tweets and retweets of a particular account, which change during the spreading phase.

\cite{ahmed2013generic} proposed 14 generic statistical features, which fall in the following four categories: interactions, posts/ tweets, URLs, and tags/mentions. These features were evaluated using NB, Jrip, and J48. The best results were achieved using J48 decision tree algorithm. \cite{Beugenilmics:2018} proposed supervised ML-based detection of organized behavior based on RF, SVM, and LR approaches. They employed user and temporal features to distinguish between three categories: organized- vs. organic- behavior, pro-Trump vs. pro-Hillary vs none and political vs. nonpolitical. Their method utilized features of collective behavior in hashtag-based tweet sets, which were collected by querying hashtags of interest. Again, the experimental results showed that RF achieves the best results with full features, however LR and SVM algorithms give better results when PCA is applied.

\cite{fernquist2018} applied RF approach to recognize automatic behaviours and detects bots that tweet about the Swedish election using meta-data and tweet (content-based) features. This approach showed better results when compared with \cite{Davis:2016}, \cite{yang2011free}, \cite{miller2014twitter} and \cite{ahmed2013generic}. They also found that the most significant feature is the number of likes the account has given divided by the number of friends the account has. The second most significant feature is the ratio between the number of followers and friends followed by the time between retweets.  \cite{al2018prediction} employed a deep regression model for sybil detection in OSNs based on 
profile-, content- and graph-based features. The system includes the following three modules: (1) data harvesting module, (2) feature extracting mechanism, and (3) deep-regression model. The system was able to achieve a good accuracy.

\cite{al2018leveraging} utilized three levels of analysis and features namely (1) user-generated content, (2) social graph connections, and (3) user profile activities, in order to detect anomalous behaviors in OSNs. The key concept in this study is leveraging contextual activity information among OSN users. They also employed principal component analysis (PCA) along with a ranking methodology to weight these features according to their relative importance in the examined dataset. Moreover, to detect a topic behavior, they used latent Dirichlet allocation (LDA) approach. Accordingly, they found that all OSN users appear to be remarkably similar until they consider their corresponding activity traits, in which significant contradictions occur. They also found that malicious users target particular topics with different activities whereas normal users involved in different activities related to several topics. Their iterative regression model achieved the best results among RF, J48, regression and SVM. 

Supervised ML approaches are unable to find zero-day malicious bots \cite{Adewole2017}. Indeed, they need a labelled dataset that captures the features and the behaviors of a diverse set of bots. To this end, \cite{Echeverria:2018} proposed “Leave-One-Botnet-Out” (LOBO), to allow supervised ML algorithms to be trained on data with multiple  types of social bots. This method is derived from cross validation in which a subset of the available data is kept out, and used for testing on $N$ number of fold. It is also worth noting that some features are computationally expensive to extract from large OSNs \cite{Adewole2017}. Interestingly \cite{cresci2015fame} showed that the best performing features are also the most costly ones. This clearly shows a trade-off between having an accurate classifier or a time-efficient one. Another drawback is that the features may be tailored for specific OSN. For instance, the number of friends request is not public on Facebook \cite{Stringhini:2010}. This suggests the need for a cross-OSN approach that takes into consideration different OSNs rather than being applicable to a particular OSN.


Apart from that, social bots evolve over time, and therefore supervised ML approaches that analyze one account at a time are unable to effectively detect this type of bots \cite{yang2013empirical,Cresci:2018,cresci2018reaction}. Moreover, such techniques may miss other stealthy bots, which are currently not being activated. Therefore, it might be useful to take into consideration a feedback from other users as a potential feature in ML approaches \cite{cao2013sybilfence}. However, this may require (1) additional data that represents possible legitimate and malicious behaviors in OSNs, (2) a model that can correctly distinguish and generalize well from the training dataset. Furthermore, \cite{chavoshi2016debot} claim that different approaches may not be able to detect all the dynamics of social bots based on limited features, which results in a smaller overlapping in terms of bot detection across these different approaches. In addition, \cite{Cresci2017} experimentally showed that most of supervised ML approaches fail by mimicking the characteristics of genuine users. Legitimate users also may purchase fake accounts to promote their profiles. This can make detection of social bots by content/behavioral based features much more challenging.

\subsubsection{Unsupervised machine learning approaches} 

Unsupervised ML approaches are able to find hidden patterns without relying on labelled data. They mainly focus on discovering specific patterns in the input. Clustering is a good example of unsupervised learning, which is used for finding useful clusters based on similar properties defined by an appropriate distance metric \cite{mypaper1}. Therefore, this approach is useful for detecting malicious campaigns, instead of inspecting individual messages of OSN users. 


\cite{Gao2012towards} proposed incremental clustering and classification approach to distinguish between spam and legitimate cluster using text-based features that fall into two categories. OSN-specific features,  and general features, which can be used to detect spam outside OSNs. The system incrementally updates the clustering result with minimal computational overhead. When the classifier detects a spam message, it will only trigger a spam alarm on that particular message, rather than on all the messages in the cluster.
However, under stealthy attacks (i.e. by reducing the attack speed and messages generated) more spam clusters become indistinguishable from legitimate message clusters.


\cite{miller2014twitter} employed modified versions of DenStream and StreamKM++ algorithms, which are based on DBSCAN and k-means, for detection of spam bots over Twitter $stream$. The original algorithms were designed to process batch data. The features used in this work fall into content- and user-based features. This approach treat the spam detection as an anomaly detection problem rather than as a classification problem. StreamKM++ achieved 99\% recall and 6.4\% false positive rate (FPR); and DenStream produced 99\% recall and 2.8\% FPR. When combined together, these algorithms reached 100\% recall and a 2.2\% FPR.

BotWalk \cite{minnich2017} 
computes an aggregated anomaly score based on an ensemble of unsupervised anomaly detection methods. BotWalk focuses on specific patterns of automated behavior based on 130 features extracted from network, content, temporal and meta-data information. Starting from a seed-bot and a set of random accounts, BotWalk retrieves each account’s details, timeline, and one-hop follower neighborhood, with the goal of maximizing the likelihood of reaching other bots across the OSN. The output of this approach is an aggregated score, obtained by combining four different anomaly detection scores. The experimental results showed that BotWalk was able to detect 7,995 previously undiscovered bots from a sample of 15 seed bots with a precision of 90\%.

\cite{Chen:2018} applied an unsupervised approach based on similarity metric. The algorithm includes two parameters minimum duplicate factor and overlap ratio, which both need to be tuned. The algorithm takes the 200 most recent tweets of each account as input and returns each bot group along with the most frequent embedded URL tweeted by that group. Then, Selenium is used to simulate a web browser and checks if a URL is malicious or not. They found that bots account for 10\% to 50\% of tweets generated from 7 URL shortening services and they were connected to large-scale spam campaigns that control thousands of domains. Briefly, we can observe that the ability to find malicious groups comes with the price of higher computational cost. Therefore, more efforts are needed to improve the speed of clustering and reduce the computational complexity. In Table 3, we provide a comparison of ML-based detection approaches.






%




\begin{ThreePartTable}
\footnotesize

\begin{TableNotes}
\footnotesize
\item GBDT: Gradient Boosting Decision Tree, XGBC: Extreme Gradient Boost Classifier.
\end{TableNotes}

\footnotesize
\begin{longtable}{| p{.12\textwidth} | p{.21\textwidth} | p{.29\textwidth} | p{.185\textwidth} | p{.075\textwidth} |} 

\caption{Comparison of machine learning-based social bot detection approaches}
\label{tab:long} \\
\hline

\makecell{Ref.} & \makecell{Approach} & \makecell{Features} & \makecell{Classes} & \makecell{Camp.\\Detection} \\
\hline
\endfirsthead

\multicolumn{5}{c}%

\endhead

\insertTableNotes
\endlastfoot

\citeauthor{Wang:2010a} & \makecell{ DT, NN, SVM, NB} & \makecell{Content- and graph-based features} & \makecell{Human or spam bot} & \makecell{\xmark} \\
\hline

\citeauthor{Chu:2012} & \makecell{Random Forest} & \makecell{Entropy of tweeting intervals as a \\measure of behavior complexity,\\tweet content and account properties} & \makecell{Human, bot or cyborg} & \makecell{\xmark} \\
\hline

\citeauthor{Stringhini:2010} & \makecell{Random Forest} & \makecell{Friend requests ratio, URL ratio,\\message similarity, friend choice,\\messages sent, friends number. } & \makecell{Spam bot \\ (Displayer, bragger,\\ poster and whisperer)\\ or not.} & \makecell{\xmark} \\
\hline

\citeauthor{Lee:2011} & \makecell{EM for cluster analysis\\Classification based on\\NB, SVM, logistic\\regression, and RF} & \makecell{User demographics, \\ user friendship networks, \\user content and user history.} & \makecell{Duplicate spammer, \\ @ spammer, malicious \\ promoter and friends \\ infiltrator or not.} & \makecell{\xmark} \\
\hline


\citeauthor{tavares:2013} & \makecell{Naive Bayes classifier} & \makecell{Distribution of Tweet time interval} & \makecell{Personal, managed\\ and bot.} & \makecell{\xmark} \\
\hline

\citeauthor{yang2013empirical} & \makecell{NB, DT, RF, DE} & \makecell{ Graph-based, neighbor-based, timing\\-based and automation-based features} & \makecell{Bot or not.} & \makecell{\xmark} \\
\hline

\citeauthor{oentaryo:2016} & \makecell{NB, RF, SVM and LR} & \makecell{ Profile-, follow-, static and \\ dynamic tweet-based features.} & \makecell{Broadcast, consumption\\ and spam bots} & \makecell{\xmark} \\
\hline

\citeauthor{Fazil:2018} & \makecell{RF, DT and \\Bayesian Network} & \makecell{ Metadata-, content-, interaction- and \\ community-based features. } & \makecell{ Automated spammer \\ or a legitimate user.} 
& \makecell{\xmark} \\
\hline

\citeauthor{Davis:2016,Onur:2017} & \makecell{Random Forest} & \makecell{User profile,friending, network,temporal, \\content and sentiment features.} & \makecell{ Bot-likelihood score \\ between [0,1].} 
& \makecell{\xmark} \\
\hline


\citeauthor{Echeverria:2018} & \makecell{Tree-based approaches: \\DT,RF,LGBM, XGBC\\and AdaBoost} & \makecell{ User and tweet-based features. } & \makecell{ Multiple bot types. }  & \makecell{\xmark} \\
\hline



\citeauthor{Beugenilmics:2018} & \makecell{RF, SVM, and LR} & \makecell{ User and temporal features.} & \makecell{Organized/organic,\\pro-Trump/-Hillary \\ political or not.} 
& \makecell{\xmark} \\
\hline

\citeauthor{fernquist2018} & \makecell{Random forest} & \makecell{Meta-data and tweet \\(content) features.} & 
\makecell{Bot or not.} 
& \makecell{\xmark} \\
\hline

\cite{al2018prediction} & \makecell{Deep regression model} & \makecell{Profile-based, content based\\and network-based.} & \makecell{Bot or not.} & \makecell{\xmark} \\
\hline

\cite{al2018leveraging} & \makecell{Iterative regression, RF,\\J48, regression \& SVM.} & \makecell{User-generated content, social graph \\connections, and user profile activities.} & \makecell{Bot or not} & \makecell{\xmark} \\
\hline



\citeauthor{Dickerson:2014} & \makecell{SVM, NB, AdaBoost, RF,\\gradient boosting, and\\extremely randomized trees} & \makecell{Tweet syntax, tweet semantics \\user behavior, network-centric\\user properties.} & \makecell{Bot or human. } & \makecell{\xmark} \\
\hline

\citeauthor{ahmed2013generic} & \makecell{NB, Jrip, and J48} & \makecell{ Interactions, posts/ tweets, URLs, and \\tags/mentions-based features.
} & \makecell{Spammer or not.} & \makecell{\xmark} \\
\hline

\citeauthor{Gao2012towards} & \makecell{Incremental clustering\\ and classification} & \makecell{Text-based features.} & \makecell{Spam or\\legitimate cluster.} & \makecell{\checkmark} \\
\hline

\citeauthor{miller2014twitter} & \makecell{DenStream and\\StreamKM++} & \makecell{content- and user-based features.} & \makecell{Bot or not.}  & \makecell{\checkmark} \\
\hline



\citeauthor{minnich2017} & \makecell{Ensemble of unsupervised\\anomaly detection\\approaches} & \makecell{Network, content, temporal\\and metadata information} & \makecell{Anomaly  score} & \makecell{\checkmark} \\
\hline

\citeauthor{Chen:2018} & \makecell{ Clustering} & \makecell{ Minimum duplicate factor \\ and overlap ratio. } & \makecell{ Bot or not.} 
& \makecell{\checkmark} \\
\hline



\end{longtable}
\end{ThreePartTable}

\normalsize

\subsection{Emerging approaches} 

\subsubsection{Detection of coordinated attacks}


Detecting the whole botnet, can help identifying a specific campaign and revealing the aggressive behavior of the botnet \cite{grimme2018changing}. Therefore, in this section we focus on approaches that aim to detect coordinated social bot attacks. In spite of the fact that sophisticated bots may use different features/characteristics, however they must have the same goal \cite{cresci2018social}, which can be inferred based on their coordinated behavior.

\cite{Fields:2018} investigated similarity and threshold-based scoring for detecting botnet campaigns. The author utilized the following features: text similarity, occurrence of similar text, summation of similarly score with $N$ neighbors, location, entropy, sentiment, user and profile-based features. These features also were assigned weights, however neither the thresholds no the assigned weights were backed by any statistical proof or explanation. 

CopyCatch \cite{Copycatch} focuses on detecting malicious Page Likes on Facebook. They search for lockstep behaviour groups of users acting together at around the same time. First, they constructed a bipartite graph between users and pages, with the time at which each edge was created. Then, they proposed subspace clustering approach in which a group of users considered suspicious if there exists a hypercube of width $2\Delta t$ in at least $m$ dimensions such that at least $n$ users fall within that hypercube where $2\Delta t$ represents the width of time window. To find suspicious clusters they define an optimization problem, in which the goal is to maximize the number of suspicious users and the number of Page Likes of suspicious users inside the appropriate cluster in the subspace. To increase the scalability, they implemented their algorithm in MapReduce framework. CopyCatch detects tightly synchronized behaviours that occur only once, however it cannot catch other generic actions such as repeatedly uploading spam-photo \cite{cao2014uncovering}. Another limitation of CopyCatch is detecting only synchronized lockstep behavior forming $blocks$, while leaving $non-overlapping$ and $partially$ $overlapping$ lockstep behaviors undetected \cite{jiang2016inferring}. These limitation are addressed in the next approaches.

In SynchroTrap \cite{cao2014uncovering} applied clustering analysis for detecting loosely synchronized actions from malicious accounts in large-scale OSNs. CopyCatch can detect fraudulent page likes that happen only once, however it cannot find other generic actions (i.e. repeatedly uploading spam-photo from the same IP addresses). Unlike CopyCatch, SynchroTrap decouples the similarity metrics from the clustering algorithm, which allows handling both once-only and other generic actions. A time-stamped action is represented with a tuple abstraction that allows the system design to be independent of the OSN applications that SynchroTrap protects. A user's actions are categorized into subsets according to the applications they belong to, which is called $application$ $contexts$. Attacker's network resource constraint is used to reduce the pairwise comparisons depending on a specific application context. Comparison results are aggregated to detect malicious actions over a longer period such as a week. For scalability concerns, they employed a parallel version of single-linkage hierarchical clustering algorithm. SynchroTrap achieved high precision (more than 99\%) and was able to find malicious activities that were undetectable by previous approaches, which also indicates that the loosely synchronized attacks have been neglected in previous defence approaches. During one month of deployment, SynchroTrap unveiled 1156 large campaigns and more than two million malicious accounts, but it does not detect them in real time. It runs on a 200-machine cluster at Facebook and it takes a few hours to process the daily data and $\sim$15 hours to process a weekly aggregation job. Being undetected for long time can cause enormous damage for OSNs.

As we mentioned previously, CopyCatch can detect only lockstep behavior forming $blocks$. To catch non-overlapping and partially overlapping lockstep behaviors \cite{jiang2016inferring} proposed LockInfer, in which they observed that lockstep behavior patterns show strange connectivity patterns, and the spectral subspace of adjacency matrix presents strange shapes in the plots. Their proposed method consists of two steps: (1) seed selection, which includes selecting nodes that behave in lockstep based on spectral-subspace plots (2) lookstep propagation, which propagates lockstep scores between followers and followees. LockInfer outperformed the existing methods with high (99–100\%) detection accuracy.

As synchronized behaviour can be viewed as an indicator of malicious activity, \cite{giatsoglou2015nd} developed a method for detecting such malicious behaviour, given a set of users and a set of retweets within a period of time. They considered timing-based features of user's retweets. Their approach consists of three steps: (1) feature subspace sweeping; (2) user scoring; (3) multivariate outlier detection. The first step includes projection of all features into all possible feature subspaces, and then segment each subspace by applying logarithmic binning, in powers of 2, in each dimension. The next step is calculating a suspiciousness score on each feature subspace. Each user is represented by a vector that contains obtained scores over all feature subspaces. The last step is multivariate outlier detection, which includes finding robust feature subspace that fits the majority of users and subsequently considers the users who are largely far from the majority as outlier users. Their method achieved 97\% accuracy and 0.82\% F1-score.

\cite{jiang2014catchsync} proposed CATCHSYNC, a graph mining approach for spotting suspicious behaviors in OSN. They considered both synchronized and rare (very different from the majority) behaviors as an indicator of suspiciousness. They chose the degree values, hubness and authoritativeness HITS score as their feature space. Specifically, they chose out-degree vs hubness, for each source node, and in-degree vs authoritativeness, for each target node. They also introduced two metrics namely synchronicity and normality to quantify these behaviors based on relative position of u’s target nodes in the feature space (i.e. indegree vs authoritativeness). Let's assume that we have a node $x$. Then, the synchronicity is measured by the average closeness between each pair of $x$’s targets. Whereas the normality is measured by the average closeness between each pair of $x$’s targets and other nodes. They showed that it is possible to find the outliers based on the existence of a provable lower bound in the synchronicity vs normality (SN)-plot. Suspicious source nodes $U_{sync}$ includes the nodes whose suspiciousness is $\alpha$=3.0 times standard deviations aways from the mean. Whereas the suspiciousness of a target node is the proportion of its sources that are reported in $U_{sync}$. The authors also showed that CATCHSYNC can restore the power law properties of the graph’s edge degree after removing the suspicious nodes from the graph. They applied CATCHSYNC on two large, real datasets 1-billion-edge Twitter social graph and 3-billion-edge Tencent Weibo social graph, and also other synthetic ones. They found that CATCHSYNC achieved better accuracy by 36\% on Twitter and 20\% on Tencent Weibo, as well as in speed when compared with \cite{perez2011spot} and \cite{outrank:2008}. However, it performs poorly when the attacks become denser due to the fact that the synchronicity will become less significant, and therefore CATCHSYNC efficient only on isolated dense blocks \cite{zhao2018actionable}. In addition, bots with few followers and followees cannot score their suspiciousness due to their poor structural information. Combining CATCHSYNC with content-based approaches can be a solution to reduce the resulted false negatives \cite{jiang2016catching}.



\cite{zhao2018actionable} proposed a method that collectively considers all target users' decisions for finding the optimal action against frauds or bots by target users themselves. Every target user can investigate the average rating of each source user who wants to contact with. If the average rating is lower than a specific threshold, then the target user assumes that the source user is more likely to be a bot. It is also expected that source users who were attacked to have a high threshold and otherwise source users will have low thresholds.

\cite{mesnards2018detecting} observed that social bots interact with humans much more frequently than they interact with each other. Based on this observation they proposed a method inspired by Ising model from statistical physics to model the network structure and bot labels. They also found that the maximum likelihood estimation of the bot labels in this model can be reduced to finding a minimum cut on an equivalent graph called \textit{energy graph}. This approach performed better than BotOrNot \cite{Davis:2016} in terms of AUC and run time. It also showed low false positives, due to the fact it takes into  consideration the interaction of the source and the target.

\cite{gupta2019malreg} proposed a method that consists of three steps: (1) community detection for coarse-grained group detection of malicious retweet groups (2) pruning algorithm for fine-grained group detection and decreasing of the number of false positive followed by supervised classifier, which makes use of a set of 23 group-based features; both entropy-based and temporal-based, to train the model. The fine-grained detection includes decomposition of an undirected weighted retweeter network for a candidate group obtained from the previous step into $n$-sub-graphs. Thereafter, they find the maximal cliques (MC) from each group such that there should be no overlap between the nodes of two maximal cliques. Then, they compute the frequency of retweets for each MC and sort them in descending order of frequency of retweets to capture cascading retweeting behaviors based on a threshold that determines the point at which there is a drastic fall in rewteeting frequency. This step divides the set into seed groups (greater than the threshold) or a set of candidate nodes (less than the threshold). Therefore, they calculate the number of common retweets between a candidate node and a seed group. A candidate node is added to the seed group with which it has maximum common retweets.
This approach achieved an accuracy of 82.88 along with an AUC of 0.921 using RF classifier.


\cite{Benigni:2019} focused on extracting largest distinct sub-graphs that exceed a predefined minimum size and density threshold based on a large sparse, weighted, reciprocal mention graph. Then, they manually inspected sub-graphs for botnet-like behavior. Sub-graph density is defined by summing link weights. In this study, they found multiple sub-communities of bots, each with a distinct intention.

\cite{Yan:2013} applied relative ranking of nodes in a graph to detect three types of botnets that attempt to hide themselves in OSNs. The first type is standalone botnets, which are isolated from the normal online social network. The second type is appendix botnets, which have only one direction following relationships. The third type is crossover botnet, which have following relationships in both directions between bots and normal users. They used the betweenness and out-degree centrality to define the core of Twitter graph. To find these types of bots the reachability is calculated at different levels : 1) for the nodes outside the core of the graph, 2) after removing strongly connected components from the graph, 3) after removing all nodes in the core from the giant weakly connected component. The nodes that exceeds a threshold at any level are monitored to find potential botnet activities. Mainly this approach decreases the number of monitored nodes except when every bot is node. The main disadvantage, however, is that it requires a clean core without malicious bots to function effectively and it has an offline component.

\cite{Gao2010} detected spam campaigns on Facebook by identifying connected subgraphs in wall posts graph. An edge is formed when two posts share the same destination URL or strong textual similarity. They assume that a) each account is limited in terms of the number of wall posts it can post, and b) messages in a single campaign are relatively bursty in time (i.e. time-synchronized). Then, they applied threshold filters based on the number of user accounts sending wall posts and time correlation within each subgraph to detect potentially malicious clusters. However, the time complexity for pairwise comparison with all wall posts is O($n^2$), which can be significant for large values of $n$. Moreover, clever bots can evade detection by gradually increasing the rate of spam messages.

\cite{lee2011content} identified top spam campaigns by applying graph mining on message graph, in which edges correspond to a content-based correlation between messages with similar \textit{talking points}. They explored three approaches for extracting campaigns: (i) loose extraction, which includes the set of all maximally connected components in the message graph, (ii) strict extraction, which includes finding maximal cliques, and (iii) cohesive extraction, which focuses on balancing loose and strict extraction by relaxing the conditions of maximal cliques. The experimental results showed that for small datasets the cohesive and strict approaches outperform the loose and cluster-based approaches. For large datasets, on the other hand, the cohesive extraction outperformed the strict extraction by combining multiple related cliques into a single campaign. In Table 4, we provide a comparison of coordinated attacks detection approaches.

\footnotesize
\begin{longtable}{| p{.1\textwidth} | p{.23\textwidth} | p{.28\textwidth} | p{.18\textwidth} | p{.08\textwidth} |} 

\caption{Comparison of coordinated attacks detection approaches}
\label{tab:long} \\
\hline

\makecell{Ref.} & \makecell{Approach} & \makecell{Features} & \makecell{Classes} & \makecell{Camp.\\Detection} \\
\hline
\endfirsthead

\multicolumn{5}{c}%
{{\bfseries \tablename\ \thetable{} -- continued from previous page}} \\
\hline 
\makecell{Ref.} & \makecell{Approach} & \makecell{Features} & \makecell{Classes} & \makecell{Camp.\\Detection} \\
\hline
\endhead

\hline
\endlastfoot

\cite{Fields:2018} & \makecell{Similarity and 
threshold-\\based scoring} & \makecell{Text similarity-features with $N$\\ 
neighbors, location, entropy, sentiment,\\ 
user and profile-based features.
} & \makecell{Bot or not} & \makecell{\checkmark} \\
\hline

\cite{Copycatch} & \makecell{Subspace clustering} & \makecell{Bipartite graph between\\users and Pages} & \makecell{ Suspicious behavior} & \makecell{\checkmark} \\
\hline

\cite{giatsoglou2015nd} & \makecell{Multivariate outlier\\
detection} & \makecell{A vector that contains obtained\\scores over all feature subspaces} & \makecell{Malicious behaviour} & \makecell{\checkmark} \\
\hline

\cite{jiang2014catchsync} & \makecell{Graph mining approach} & \makecell{Social graph} & \makecell{Suspicious behavior} & \makecell{\checkmark} \\
\hline

\cite{cao2014uncovering} & \makecell{Single-linkage \\hierarchical clustering} & \makecell{Generic time-stamped actions} & \makecell{Bot or not} & \makecell{\checkmark} \\
\hline

\cite{jiang2016inferring} & \makecell{Graph-based lockstep \\behavior inference} & \makecell{Adjacency matrix and\\spectral
subspaces} & \makecell{Lockstep behaviors} & \makecell{\checkmark} \\
\hline

\cite{zhao2018actionable} & \makecell{ Finding optimal threshold\\for each target user } & \makecell{Adjacency matrix} & \makecell{User’s blocklist\\thresholds} & \makecell{\checkmark} \\
\hline

\cite{mesnards2018detecting} & \makecell{Ising model} & \makecell{Energy graph mapped from an \\interaction network/graph} & \makecell{Bot or not} & \makecell{\checkmark} \\
\hline

\cite{gupta2019malreg} & \makecell{Three-step approach:\\1- community detection\\2- pruning algorithm\\3- supervised classification.} & \makecell{Social graph, group-based features:\\ (1) entropy- and (2) temporal-based.} & \makecell{Malicious or benign\\ group} & \makecell{\checkmark} \\
\hline

\cite{Benigni:2019} & \makecell{Extracting largest\\distinct subgraphs} & \makecell{Large sparse, weighted, \\reciprocal mention graph} & \makecell{Promoted account, \\community influencer. }  & \makecell{\checkmark} \\
\hline

\cite{Yan:2013} & \makecell{ Graph-based approach } & \makecell{ Original twitter graph } & \makecell{ Standalone, appendix \\and crossover botnets.} 
& \makecell{\checkmark} \\
\hline

\cite{Gao2010}& \makecell{Detecting connected\\subgraphs.} & \makecell{Wall posts graph formed according\\to same destination URL or strong\\textual similarity.} & \makecell{Malicious (URL/Post)\\or not.} 
& \makecell{\checkmark} \\
\hline

\cite{lee2011content} & \makecell{Graph mining} & \makecell{ Message graph, in which edges\\correspond to a content-based\\correlation between messages.} & \makecell{Top campaigns.} 
& \makecell{\checkmark} \\
\hline

\end{longtable}

\normalsize

\subsubsection{Other emerging approaches}

As malicious spam bots cannot control their followers to follow them, a lot of legitimate OSN users but few of them follow back these malicious spammers, taking this observation into consideration \cite{feng2017groupfound} proposed GroupFound, a method that exploits the bi-follow relationship of a target-node to establish an undirected graph. First, they get the two layers of neighbor nodes of the target-node. Then, they compute the similarity of every two nodes among the neighbor-nodes of a target node by computing the Jaccard index. Then, they obtain a similarity matrix, which will be used to calculate the groups of target-node. Each row is considered as a record and every column can be viewed as a feature. For each row they calculate a mean value vector and the distance from each node of the neighbor-nodes of the target node to the mean. The algorithm is mainly based on hierarchical clustering. GroupFound achieved a detection rate of 86:27\% with a  false positive rate of 8:54\%.

\cite{cresci2018social} observed that it is not enough to merely depend on the history of previous behaviour records to detect new generations of spam bots. Instead, we need to investigate collective behaviors of users' groups to determine whether these account are bots or not. To this end, they proposed DNA sequences for modeling the behaviors of OSN users. Their model considers sequences as ordered lists of symbols, with variable length, taken from a relatively small alphabet. The longest common sub-string (LCS) between two or more DNA sequences can be used to measure similarities between these sequences. They found that the LCS of spam bots are long even when the number of accounts grows, whereas legitimate users show low to zero similarities. Accordingly they were able to uncover traces of an automated and synchronized activity. Based on LCSs, they proposed supervised and unsupervised approaches where the supervised approach achieves slightly better results. The experimental results showed that this approach was able to outperform \citeauthor{yang2013empirical}, \cite{ahmed2013generic} and \cite{miller2014twitter}. Interestingly, the very low recall of \cite{yang2013empirical} can be seen as an evidence of a new generation of social bots that are hard to detect when they are considered individually even using current state-of-the-art algorithms. \cite{cresci2018social} approach is flexible in terms of not focusing on specific characteristics. Moreover, it reduces the cost for data gathering by not considering the properties of the social graph. However, to detect more sophisticated types of bots, partial matches can be considered instead of exact ones. In other words, instead of using the longest common sub-string, it is possible utilize longest common sub-sequence metric as they suggested.









\cite{clark2016sifting} focused on three distinct classes of automated tweeting: robots, cyborgs and human spammers. Their method classifies accounts merely based on linguistic attributes. They make use of the following three distinct linguistic features: (i) average pairwise tweet dissimilarity (ii) word introduction rate decay parameter and (iii) average number of URLs per tweet. They found that for legitimate users, these three attributes are densely clustered, but they can vary greatly for automatons. Their approach classifies each user as an automated account if their feature falls further than $n$ standard deviations away from the legitimate user mean, for varying $n$. This step is referred to as calibration phase. Moreover, they found many cyborgs send incomplete messages followed by an ellipsis and a URL. The experimental results showed that the AUC increases as the number of collected tweets increase. However, it showed a large false positive rate (22\%) for human accounts when the same approach was applied to another set of accounts collected from social Honeypot experiment \citeauthor{Lee:2011}. They achieved better results by applying another calibration phase.


\cite{varol2017early} observed that promoted trends on Twitter are sustained by organic activity before promotion, and therefore they are essentially indistinguishable from organic ones until the promotion triggers the trending behaviour. Consequently as more users join the conversation, these trends tend to be more indistinguishable from the organic ones.
Therefore, they focused on early detection of promoted trending memes based on the temporal sequence of messages associated with a particular trending hashtag and classifying it as either promoted or organic. They used 487 features, which can be categorized into five classes: network structure and information diffusion patterns, content and language, sentiment, timing, and user meta-data. They also applied wrapper approach to select features where they train and evaluate models using the candidate subsets of features and expand the set of selected features using greedy forward feature selection. A k-nearest neighbor approach with Dynamic Time Warping (kNN-DTW) was proposed to deal with multi-dimensional time series classification problem. Time series for each feature were processed in parallel using dynamic time warping (DTW), which measures the similarity between two time series after finding an optimal match between them by $warping$ the time axis, which, in turn, allows capturing some non-linear variations in time series. They also found that content- and user-based features are the most useful features for early detection of the promoted content whereas timing- and network-based features become important when more users involve after trending. Overall, their approach showed 75\% AUC score for early detection, increasing to above 95\% after trending.

\cite{el2018supervised} introduced a hybrid approach, in which graph-based method is coupled with machine learning classifiers in a probabilistic graphical model framework. They choose to use a graph based on the similarity between users’ applications, rather than using the social structure of the network. They build a Markov Random Field  (MRF) model on top of the resulting similarity graph. They used supervised classifiers to calculate the initial beliefs of MRF. On the other hand, the similarity between users is used to propagate beliefs about their labels. The graphical inference phase only alters beliefs on connected nodes, which is efficient against social bots where the presence groups of connected spammers is common. They applied joint optimization using Loopy Belief Propagation over the MRF to obtain the most accurate labels, which permits to correct misclassified labels from baseline supervised classifiers. They used profile-, network-, content- and behavioral-based features. Their novel features and approach successfully increased both precision and recall even for off-the-shelf classifiers when compared with the features used in \cite{benevenuto2010detecting} and \cite{stringhini2012poultry}. Moreover, their hybrid approach enhanced the overall accuracy from 0.918\% to 0.952\% when compared to SVM alone where they used the same features for these two approaches. However, as we mentioned above, the graphical inference phase alters beliefs only on connected nodes. Therefore, when spam nodes are isolated from each other, the performance of this approach will be equivalent to a conventional supervised classifier. 

\cite{lee2014early} proposed a method to detect potentially malicious account groups around their creation time based on the differences between automatically generated account names and human-made ones. An agglomerative hierarchical clustering approach is applied to group accounts sharing relevant name-based features within a short period of time. Then, they classify malicious account clusters using SVM classifier. Their approach achieved 1.98\% FNR and 20.74\% FPR. The goal here is to notify back-end detection systems to monitor these account groups, and consequently take further action them. The main drawback of this approach is that it depends on the characteristics of malicious account names. Therefore, attackers may design advanced name generation techniques to evade detection. The worst case may happen when the attackers mimic the name-based features of legitimate users.






\cite{chavoshi2016debot} proposed DeBot, a dynamic-time-warping (DTW)-based correlation approach to find bots with high temporally correlated activities. They developed a novel lag-sensitive hashing technique to group correlated users based on warping
correlations. DTW is applied in order to compute the distance between two users. DeBot consists of the following four stages: The first stage includes collecting tweets that contain selected keywords for a per-defined period, then it forms the time series of activities at every second for all of the user-accounts and filters out users with just one activity. These series are processed by taking the activity time series of all the users as input and hashes each of them into multiple hash buckets to detect correlated activity patterns between two or multiple users by reporting sets of suspicious users that collide at the same hash buckets. The third stage includes monitoring the activity of suspected users via the stream API. The last step includes single-linkage hierarchical clustering on the pairwise DTW distances calculated from suspicious time series. DeBot detects less number of bots than BotorNot, due to the fact that BotorNot is trained based on English-language tweets, while DeBot catches all languages just based on temporal synchronicity. The authors also claim that other approaches may not be able to detect all the dynamics of social bots based on limited features, which results in a smaller overlapping between these different approaches. DeBot was able to detect 544,868 unique bots through one year period. A smart attacker, however, can evade detection by inserting unbounded random time delays among the same tweet from many accounts \cite{chavoshi2016identifying}.

\cite{Perna:2018} introduced Learn To Rank (LTR), a supervised approach to learn a ranking model from annotated instances to detect social bots at different severity levels. LTR learns from training data, which corresponds to multiple queries, whose annotations are according to degree of relevance of objects to a given query. To this end, they extracted three groups of features: (1) static (user-based, network-based, temporal-, content-based), (2) aggregate- and (3) query(keyword)-based features where the feature extraction step was performed via Twitter API and services of BotOrNot, BotWalk and DeBot approaches. For relevance labeling, they use three approaches: (1) binary relevant balanced (BB) selection (2) binary relevance/unbalanced (BU) selection, and (3) balanced (Grad) selection with 7 grades of bot status. For feature selection, they used principal component analysis (PCA), information gain (IG) and gain ratio (GR) attribute evaluation, OneR, correlation-based feature selection (CFS), and learner-based feature selection with J48. Content-based and aggregate features were mostly selected by the above methods. RankNet \cite{RankNet} ,Coordinate Ascent \cite{CoordinateAscent}, AdaRank \cite{AdaRank} and LambdaMART \cite{LambdaMART} as ranking models. The most successful models were Coordinate Ascent and LambdaMART under different scenarios. The aggregate features followed by content-based features, reveal to be a more robust to all methods. Interestingly, LambdaMART and Coordinate Ascent were able to achieve very good ranking prediction accuracy when trained on the subspace of aggregate features, which again explains the importance of aggregate features. Under the BB setting, the methods achieved low variance in their performance. However, under UB setting they showed a high variance. Under Grad setting, Coordinate Ascent and LambdaMART were more robust w.r.t. the assessment criteria.

%






\cite{morstatter2016new} proposed BoostOR, a method that optimizes the F1 score through boosting. Their method gives higher weights for mislabeled bots while downweights mislabeled legitimate users. In addition, they employed latent Dirichlet allocation (LDA) to obtain a topic representation of each user, which can be viewed as a probability distribution over $k$ topics. The experimental results on two real-world datasets showed that their method achieved better results over SVM, AdaBoost, as well as other common heuristics used to detect bot accounts. Table 5, provides a comparison of emerging detection approaches.

\footnotesize
\begin{longtable}{| p{.1\textwidth} | p{.21\textwidth} | p{.30\textwidth} | p{.18\textwidth} | p{.08\textwidth} |} 

\caption{Comparison of emerging detection approaches}
\label{tab:long} \\
\hline

\makecell{Ref.} & \makecell{Approach} & \makecell{Features} & \makecell{Classes} & \makecell{Camp.\\Detection} \\
\hline
\endfirsthead

\multicolumn{5}{c}%
{{\bfseries \tablename\ \thetable{} -- continued from previous page}} \\
\hline 
\makecell{Ref.} & \makecell{Approach} & \makecell{Features} & \makecell{Classes} & \makecell{Camp.\\Detection} \\
\hline
\endhead

\hline \multicolumn{5}{|r|}{{Continued on next page}} \\ \hline
\endfoot

\hline
\endlastfoot

\cite{feng2017groupfound} & \makecell{Hierarchical clustering
} & \makecell{Local graph (i.e. two layers of\\neighbor nodes of the target-node)} & \makecell{Spam or normal\\account} & \makecell{\xmark} \\
\hline

\cite{cresci2018social}  & \makecell{Digital DNA-based \\approach.} & \makecell{Behavior-based features.} & \makecell{Bot or not.} & \makecell{\checkmark} \\ \hline



\cite{clark2016sifting} & \makecell{Natural language approach} & \makecell{Average pairwise tweet dis-\\similarity. Word introduction\\ rate decay parameter. Average\\number of URLs per tweet.} & \makecell{Robots, cyborgs\\ \& human spammers.} & \makecell{\xmark} \\
\hline


\cite{varol2017early} & \makecell{k-NN with dynamic\\time warping} & \makecell{Network structure and information\\diffusion patterns, content and language,\\sentiment, timing, and user meta-data} & \makecell{Organic and promoted\\trends.} & \makecell{\checkmark} \\
\hline

\cite{el2018supervised} & \makecell{MRF with initiated\\beliefs using SVM} & \makecell{Profile, network, content and\\ behavioral features} & \makecell{Verified users (cyborg),\\human users, trend\\hijackers, promotional\\spambots.} & \makecell{\xmark} \\
\hline

\cite{lee2014early} & \makecell{Clustering \& classifying\\malicious groups.} & \makecell{Account name-based features.} & \makecell{Legitimate or\\malicious clusters.} & \makecell{\checkmark} \\
\hline

\cite{morstatter2016new} & \makecell{BoostOR} & \makecell{A topic representation of each user\\ based on LDA.} & \makecell{Bot or not} & \makecell{\xmark} \\
\hline

\cite{chavoshi2016debot} & \makecell{
Single-linkage\\hierarchical clustering} & \makecell{DTW distances calculated\\from suspicious time series.} & \makecell{Bot or not} & \makecell{\checkmark} \\
\hline

\cite{Perna:2018} & \makecell{
RankNet, Coordinate\\Ascent, AdaRank\\and LambdaMART.} & \makecell{Three groups of features: (1) static\\(user-based,network-based, temporal,\\content-based),(2) aggregate and \\ (3) query(keyword)-based features.} & \makecell{ Ranking bots } & \makecell{\xmark} \\
\hline

\end{longtable}

\normalsize

\section{Datasets and Findings}
The lack of public ground-truth data is considered to be the main challenge that hinders appropriate evaluation of social bot detection approaches. Comparing detection approaches is subjected to many factors such as dataset size, the number of features considered, the ground-truth quality, data crawling process, the type of method adopted \cite{Adewole2017}. Honey-profiles can also be used for attracting malicious social bots and identifying their different activities and behaviours. Researchers mainly create fake accounts and record the interactions with these accounts. Since these accounts are fake and generally inactive, it is assumed that these interactions can only come from malicious accounts \cite{Echeverria:2018}. Honey profiles can be designed to look legitimate, while other honeypot profiles may explicitly indicate that they are not real. They may also attract fake profiles by means of paid campaigns \cite{de2014paying}. Honeypot datasets were used widely in different studies \cite{Stringhini:2010,Lee:2011,Subrahmanian:2016,Davis:2016,Echeverria:2018}. Tables (6,7,8) provide a summary of datasets and corresponding findings of the previously studied methods. It is worth mentioning that we did not include details related to datasets used in graph-based approaches as they already were provided in \cite{ramalingam2018fake}.


\footnotesize
\begin{ThreePartTable}
\begin{TableNotes}
\footnotesize
FR: Friend Request, M: Message, DR: Detection Rate, FPR: False Positive Rate, TPR: True Positive Rate.
\end{TableNotes}

\begin{longtable}{| p{.12\textwidth} | p{.075\textwidth} | p{.325\textwidth} | p{.355\textwidth}|} 

\caption{Comparison of datasets used in machine learning-based detection approaches and their related findings}
\label{tab:long} \\
\hline

\makecell{Ref.} & \makecell{Honeypot} & \makecell{Dataset} & \makecell{Findings}  \\
\hline
\endfirsthead

\multicolumn{4}{c}%
{{\bfseries \tablename\ \thetable{} -- continued from previous page}} \\
\hline 
\makecell{Ref.} & \makecell{Honeypot} & \makecell{Dataset} & \makecell{Findings} \\
\hline
\endhead

\hline \multicolumn{4}{|r|}{{Continued on next page}} \\ \hline
\endfoot

\hline
\endlastfoot

\citeauthor{Wang:2010a} & \makecell{\xmark} & \makecell{Collected from Jan. 3 to Jan 24, 2010.\\ Totally 25,847 users \& around 500K tweets.
} & \makecell{Naive Bayes achieved the best results:\\precision, recall \& F-measure of 0.917.}  \\
\hline

\citeauthor{Chu:2012} & \makecell{\xmark} & \makecell{Collected 500,000 Twitter users with\\ more than 40 million tweets.} & \makecell{Average TPR of 96\%.} 
\\
\hline

\citeauthor{Stringhini:2010} & \makecell{\checkmark} & \makecell{300 honey profiles were created, 100 for\\each social network. Collected from\\June, 2009 to June, 2010. Total 4250\\friend requests and 85,569 messages.} & \makecell{Facebook: 4.51\% FR. - M-5.35\%. \\
MySpace: 36.3\% FR. - M-0\% \\
Twitter: 90.93\% FR. - M-86.4\%} \\
\hline

\citeauthor{Lee:2011} & \makecell{\checkmark} & 
\makecell{60 honey profiles. Total of 36,000 spammers.\\Collected from 30 Dec,2009 to 2 Aug, 2010.} & \makecell{Boosting of Random Forest achieved\\ the best results (Accuracy of 98.62\%).}   \\
\hline


\citeauthor{tavares:2013} & \makecell{\xmark} & \makecell{Collected dataset contains 244\\ acounts with total 164975 tweets.
} & \makecell{Accuracy of 84.6\% when classifying personal and\\managed accounts. 75.8\% when classifying\\personal, managed and bot accounts.
}   \\
\hline

\citeauthor{yang2013empirical} & \makecell{\xmark} & \makecell{Two different datasets. The first dataset\\consists of 20K benign account, selected\\from our crawled 500. The second\\dataset includes 35K Twitter accounts\\randomly selected 3,500 accounts from\\a dataset that contains a total of 500K \\accounts, 2060 of which were bots.} & \makecell{RF achieved false positive rate of 0.4\%,\\detection rate of 84.8\% and F1-score of 90\%} \\
\hline

\citeauthor{oentaryo:2016} & \makecell{\xmark} & 
\makecell{Generated by users in Singapore \& collected\\from 1 Jan. to 30 Apr. 2014. It includes\\a total of 159,724 accounts with 589 bots.} & \makecell{LR and SVM showed the best results.}   \\
\hline

\citeauthor{Fazil:2018} & \makecell{\xmark} &  \makecell{They used 1KS-10KN, a dataset that\\ contains 11000 labeled users\\(10K benign users \& 1K spammers)\\ with total of 1354618 tweets.}  &  \makecell{ Outperformed \cite{yang2013empirical}.\\
RF achieved the best results in\\
terms of DR, FPR, and F-Score.
} \\
\hline

\citeauthor{Davis:2016,Onur:2017} & \makecell{\checkmark} & \makecell{A dataset of 15K manually verified bots and\\16k account with more than 5.6M tweets.} & \makecell{0.95 AUC (Area Under ROC Curve).}   \\
\hline

\citeauthor{Echeverria:2018} & \makecell{\checkmark} & \makecell{ A botnet dataset that contains aggregated\\content generated from a variety of bot\\datasets and a second legitmate user\\dataset. The final aggregated bot\\dataset contains over 1.5M bots\\ with all their available tweets.} & \makecell{Light Gradient Boosting Machine (LGBM)\\ achieved an accuracy of 97\%.} \\
\hline


\citeauthor{Beugenilmics:2018} & \makecell{\xmark} & \makecell{Analysis of over 200 million tweets which\\ were mainly posted during the 2016 US\\presidential election. The model is trained\\using a training data set with 851 records.} & \makecell{RF showed high scores with full features, with\\an average accuracy and f-score greater than\\0.95, while LR and SVM showed better results\\when PCA is applied.} 
\\
\hline

\citeauthor{fernquist2018} & \makecell{\xmark} & \makecell{They used three datasets for training their \\model. The first one was crawled during Oct.\\and Nov. 2015 and contains 647 bots and\\1367 genuine accounts. The second dataset\\consists of 591 bots and 1,680 genuine\\accounts. The third dataset is manually\\annotated and consists of 519 human\\ accounts and 355 bot accounts. The test\\dataset consists of 991 social spam bots\\and 991 genuine accounts.
\\ } &  \makecell{Their model achieved the best results when\\compared to other supervised models. However,\\an unsupervised approach \cite{cresci2016dna}\\outperformed their model in terms of accuracy,\\precision and F1-score. They achieved\\the highest recall among all other models.
} \\
\hline

\cite{al2018prediction} & \makecell{\xmark} & \makecell{ Two large datasets included 25,510 and 2200\\manually annotated honest and sybil accounts,\\along with 13,957 and 940 profiles’ accounts.\\They were generated from posts related to\\the USA Election 2016. } & \makecell{An accuracy of 86\% when fed with noisy data. }
\\
\hline

\cite{al2018leveraging} & \makecell{\xmark} & \makecell{609M tweets from nearly 364K individual\\ Twitter accounts, as well as 51K YouTube\\users
with 13M channel activities.} & \makecell{In terms of F-measure, RF and iterative regression\\ achieved 96\%. J48 achieved 94\%. Regression\\and SVM achieved 95\% and 84\% respectively.}  \\
\hline

\citeauthor{Dickerson:2014} & \makecell{\xmark} & \makecell{Indian Election Dataset (IEDS) collected \\from July 15, 2013 to March 24 2014.\\It contains 7.7M tweets by over 555K users.} & \makecell{
AUC of 0.73.
}   \\
\hline

\citeauthor{ahmed2013generic} & \makecell{\xmark} & \makecell{Facebook dataset contains 320 profiles,\\165 spam and 155 normal user profiles.\\Twitter dataset contains 305 profiles,\\160 spam and 145 normal user profiles. } & \makecell{The results obtained on a combined dataset\\has DR as 0.957 and FPR as 0.048,\\whereas on Facebook dataset the DR and\\FPR values are 0.964 and 0.089 respectively\\and on Twitter dataset the DR and FPR\\values are 0.976 and 0.075 respectively.}
\\
\hline

\citeauthor{Gao2012towards} & \makecell{\xmark} & \makecell{ This study contains two datasets: Facebook\\data is the same as their previous study.\\ Twitter set contains over 17M tweets related\\to trending topics that were generated\\between Jun. 1 and Jul. 21, 2011.} & \makecell{TPR of 80.9\% and FPR of 0.19\%.} 
 \\
\hline

\citeauthor{miller2014twitter} & \makecell{\xmark} & \makecell{The dataset for this study included 3239\\user accounts with a sample tweet from \\each account. 208 spam accounts and 3031\\randomly selected verified normal users.} & \makecell{StreamKM++ achieved 99\% recall and a 6.4\%\\FPR;and DenStream produced 99\% recall and\\a 2.8\% FPR. When combined together, these\\algorithms reached 100\% recall and a 2.2\% FPR.
}
\\
\hline

\citeauthor{minnich2017} & \makecell{\xmark} & \makecell{A dataset that contains 362K nodes\\with 226M edges\\}  & \makecell{Detected 7,995 previously undiscovered bots\\from a sample of 15 seed bots with a\\precision of 90\%}  \\
\hline



\citeauthor{Chen:2018} & \makecell{\xmark} &
\makecell{A white-list of popular and trustworthy \\URLs that includes nine most widely\\used Twitter URL shortening services,\\collected 50,000 tweets from each service.\\They collect most recent 200 tweets\\from every account in each group.} & \makecell{Bots account for 10\% to 50\% of tweets\\generated from 7 URL shortening services. \\They also found that bots using shortened\\URLs are connected to large-scale spam\\campaigns that control thousands of domains.}   \\
\hline

\end{longtable}
\end{ThreePartTable}

\footnotesize
\begin{ThreePartTable}

\begin{longtable}{| p{.13\textwidth} | p{.075\textwidth} | p{.36\textwidth} | p{.34\textwidth}|} 

\caption{Comparison of datasets used in coordinated attacks approaches and their related findings}
\label{tab:long} \\
\hline

\makecell{Ref.} & \makecell{Honeypot} & \makecell{Dataset} & \makecell{Findings}  \\
\hline
\endfirsthead

\multicolumn{4}{c}%
{{\bfseries \tablename\ \thetable{} -- continued from previous page}} \\
\hline 
\makecell{Ref.} & \makecell{Honeypot} & \makecell{Dataset} & \makecell{Findings} \\
\hline
\endhead

\hline \multicolumn{4}{|r|}{{Continued on next page}} \\ \hline
\endfoot

\hline
\endlastfoot

\cite{Fields:2018} & \makecell{\xmark} & \makecell{ Several datasets that contained\\approximately 933.222 tweets } & \makecell{
Detected 14,585 bots. Under 11,000\\of these were unique accounts.}  \\
\hline

\cite{Copycatch} & \makecell{\xmark} & \makecell{Facebook dataset contains 3.3 billion Likes\\and synthetic data that consists of a bipartite\\graph between 38 million and 10 million\\nodes with 410 million edges \\} & \makecell{Fast detection with high accuracy\\and low false positive rate.}  \\
\hline

\cite{giatsoglou2015nd} & \makecell{\xmark} & \makecell{
12M retweets to posts of 298 active Twitter users \\
} 
& \makecell{97\% accuracy and 0.82 F1-score. }  \\
\hline

\cite{jiang2014catchsync} & \makecell{\xmark} & \makecell{ Three sets of synthetic data that
contains\\approximately 1M, 2M, or 3M nodes. In addition\\to two real
world datasets: 1) Twitter dataset\\that
contains 41,652,230 nodes with 
1,468,365,182\\edges from Titter
(July 2009), and 2) Weibo\\dataset
which contains 117,288,075 node
with\\3,134,074,580 edges (Jan 2011).\\
} & \makecell{Achieved better accuracy by 36\% on\\Twitter and 20\% on Tencent Weibo, as\\well as in speed when compared with \\ \cite{perez2011spot}}  \\
\hline

\cite{cao2014uncovering} & \makecell{\xmark} & \makecell{ One-month execution log\\at Facebook in August 2013 } &\makecell{Achieved more than 99\% precision. Finds\\malicious activities that were undetectable\\by previous approaches, unveiled 1156 large\\campaigns and more than two million\\malicious accounts.}  \\
\hline

\cite{jiang2016inferring} & \makecell{\xmark} & \makecell{ Actor-movie IMDb dataset, US patent\\citation network with 3,774,768 vertices\\ \& 16,518,947 edges, \& a synthetic dataset. \\ } & \makecell{Outperforms the existing methods with\\99–100\% detection.}  \\
\hline

\cite{zhao2018actionable}  & \makecell{\xmark} & \makecell{
Synthetic dataset with 2K sources and 
2K targets.\\In addition to product 
review dataset from\\Amazon(2015) with 4,552 users \& 6,347 products \\
} & \makecell{99\% accuracy and 99.8\% F1 score}  \\
\hline

\cite{mesnards2018detecting} & \makecell{\xmark} & \makecell{
Twitter data from six different events \\
1- Pizzagate (Nov-Dec 2016) which\\contains 1025911 tweets - 176822 users  \\
2- BLM 2015 (Jan-Dec 2015) which\\contains 477344 tweets - 242164 users  \\
3- US election (Sep-Oct 2016) which\\contains 2435886 tweets - 995918 users \\
4- Macron leaks (May 2017) which\\contains 570295 tweets - 150848 users  \\
5- Hungary election (Apr 2018) which\\contains 504170 tweets - 198433 users  \\
6- BLM 2016 (Sep 2016) which\\contains 1274596 tweets 545937 users.\\Approximately 10\% of the accounts\\were labeled as bots. \\
} 
& \makecell{Outperformed BotOrNot \citeauthor{Davis:2016}\\in terms of AUC, runtime\\and false positive rate. }  \\
\hline

\cite{gupta2019malreg} & \makecell{\xmark} & \makecell{
Three datesets related to different political\\ events: 1- UK General Election (2017)\\ with 1,459,205 reweets \& 443,913 users\\
2- Indian banknote demonetization (2016)\\ with 2,015,101 retweets \& 288,487 users\\
3- Delhi Legislative Assembly Election (2013)\\with 6,800,687 tweets and 297,793 users.
} 
& \makecell{82.88\% accuracy and an AUC of 0.921 \\using Random Forest classifier.}  \\
\hline

\cite{Benigni:2019} & \makecell{\xmark} & \makecell{Includes 3 datasets. The first one has 106K users\\and 268 million tweets. The 2nd one has 92,706\\users and 212 million tweets. The 3rd dataset\\has 87,046 users and 179 million tweets.} & 
\makecell{ Multiple sub-communities of bots\\ or cyborgs within each dataset.}   \\
\hline

\cite{Yan:2013} & \makecell{\xmark} & \makecell{Collected between Jun. 31 and\\
Sep. 24 in 2009. It contains\\
41,652,230 user profiles and\\
1,838,934,111 tweets.} & 
\makecell{The main advantages is small number of\\monitored nodes except when every bot is\\ node. The main disadvantage is that it\\ requires a clean core without malicious\\bots to function effectively and it has\\an offline component.}   \\
\hline

\cite{Gao2010} & \makecell{\xmark} & \makecell{Facebook dataset contains 187M\\posts generated by roughly 3.5M \\
users in between Jan. of 2008 and \\
Jun. of 2009.} & \makecell{ TPR of 96.1\% and 93.9\% for malicious URLs\\and posts respectively. FPR of 3.9\% \& 6.1\% \\ for benign URLs and wall posts respectively.\\ They found 200,000 malicious wall posts with \\embedded URLs, obtained from more than\\57,000 malicious accounts, of which 97\%\\ found to be compromised accounts. } \\
\hline

\cite{lee2011content} & \makecell{\xmark} &  \makecell{It contains two datasets. The first\\dataset contains 1912 tweets. The\\second one contains $\sim$1.5M posts\\between 1 and 7 Oct. 2010.} & 
\makecell{For the first (small) dataset the cohesive and\\strict approaches outperformed the loose\\and cluster-based approaches. For the other\\(large) dataset cohesive campaign detection\\ approach outperformed the strict campaign\\detection by combining multiple related\\cliques into a single campaign. 
The largest\\campaign contains 560
vertices and\\it is a spam campaign.
} \\
\hline

\end{longtable}
\end{ThreePartTable}

\footnotesize
\begin{ThreePartTable}

\begin{longtable}{| p{.14\textwidth} | p{.075\textwidth} | p{.355\textwidth} | p{.33\textwidth}|} 

\caption{Comparison of datasets used in emerging detection approaches and their related findings}
\label{tab:long} \\
\hline

\makecell{Ref.} & \makecell{Honeypot} & \makecell{Dataset} & \makecell{Findings}  \\
\hline
\endfirsthead

\multicolumn{4}{c}%
{{\bfseries \tablename\ \thetable{} -- continued from previous page}} \\
\hline 
\makecell{Ref.} & \makecell{Honeypot} & \makecell{Dataset} & \makecell{Findings} \\
\hline
\endhead

\hline \multicolumn{4}{|r|}{{Continued on next page}} \\ \hline
\endfoot

\hline
\endlastfoot

\cite{feng2017groupfound} & \makecell{\xmark} & \makecell{ Sina Weibo dataset, collected in Jan. 2015\\consists of 246,898 accounts, 224,481 normal\\accounts and 22,417 spam accounts.\\ } & \makecell{A detection rate of 86:27\% with a false\\positive rate of 8:54\%. It also outperformed\\PageRank and SybilDefender.}  \\
\hline

\cite{cresci2018social} & \makecell{\xmark} & \makecell{It included the following datasets: the first one\\included 991 spambots with 1610176 tweets\\related to Mayoral election in Rome, in 2014.\\The other one included 464 spambots with \\1418626 tweets try to advertise a subset of\\ products on sale on Amazon. The last one\\is a human dataset, which included 3474\\account with 8377522 tweets. } & \makecell{ Outperformed \citeauthor{yang2013empirical},\\ \cite{ahmed2013generic} and\\ \cite{miller2014twitter}. } \\  \hline



\cite{clark2016sifting} & \makecell{\checkmark} & \makecell{1\% sample of Twitter’s streaming API
containing\\geospatial metadata between Apr - Jul 2014\\and it includes the most active 1000 users.\\ It also included another set of accounts\\collected from social honeypot experiment.\\
} & \makecell{Precision, recall \& F-measure of 0.917.}  \\
\hline


\cite{varol2017early} & \makecell{\xmark} & \makecell{The promoted trend dataset was collected\\between 1 Jan and 31 Apr 2013 with 2385\\tweets and 2090 unique user. The organic \\dataset was collected between 1-15 Mar 2013 \\with 3692 tweets and 2828 unique users. }  & \makecell{An AUC of 75\% for early detection,\\which also increased above 95\%\\ after trending.}  \\
\hline

\cite{el2018supervised} & \makecell{\xmark} & \makecell{Collected between 5 and 21 Oct 2017.\\It contains a random sample of 20M \\tweets from 12M active users.} & \makecell{Enhanced the overall accuracy from\\0.918 to 0.952 when compared to SVM.} \\
\hline

\cite{lee2014early} & \makecell{\xmark} & \makecell{It consists of 4,687,345 Twitter accounts\\created between April 2011 and October\\2011 among 18,289 of them were verified\\ Twitter accounts. } & \makecell{Achieved 1.98\% FNR and 20.74\% FPR.}  \\
\hline

\cite{morstatter2016new}  & \makecell{\checkmark} & \makecell{It contains two datasets. The first one is related to\\Arab Spring in Libya and was  collected\\between 3rd Feb, 2011 to 21st Feb, 2013.\\It contains 94535 unique users, 7.5\% of them\\are bots. The second one is arabic honeypot\\dataset in which, they collected 6285 unique,\\accounts 3602 of them were active bots.} 
& \makecell{Outperformed SVM, AdaBoost\\as well as other common heuristics\\used to detect bot accounts.}  \\
\hline

\cite{chavoshi2016debot} & \makecell{\xmark} & \makecell{The results were relatively supported\\by OSNs (Twitter) and other approaches.} & \makecell{Achieved 94\% precision and detected \\544,868 unique bots through one\\year period.
}  \\
\hline

\cite{Perna:2018} & \makecell{\checkmark} & \makecell{ A total of about 19K accounts,\\11K of which correspond to bots\\and the remaining are non-bot accounts\\from datasets used in \citeauthor{Cresci2017}\\ \citeauthor{gilani2017f,morstatter2016new} \\and \citeauthor{Onur:2017}. }  & \makecell{The most successful models were\\ Coordinate Ascent and LambdaMART.\\ Under the BB setting the methods showed\\low variance in their performance but\\under UB setting they had high variance.\\For Grad setting, Coordinate Ascent and\\LambdaMART were more robust\\w.r.t. the assessment criteria. }  \\
\hline

\end{longtable}
\end{ThreePartTable}

\normalsize

\section{Conclusion}
This study reviewed the state of the art of malicious social bots in terms of their stealthy behavior and detection approaches. We meticulously investigated three main network-based detection approaches namely, machine learning, graph-based and emerging approaches. As a result, we proposed a refined taxonomy that can be leveraged by OSN administrators and researchers to improve the current defence strategies against malicious social bots. In addition, datasets used and their corresponding findings were studied in detail.

\section*{Acknowledgement}
We would like to thank Gonca Gürsun and Huseyin Ulusoy for their fruitful comments on the manuscript, which helped us improve the quality of this work.


\begin{thebibliography}{}

\bibitem[\protect\astroncite{Abokhodair et~al.}{2015}]{Abokhodair:2015}
Abokhodair, N., D.~Yoo, and D.~W. McDonald\leavevmode\nopagebreak\newline 2015.
\newblock Dissecting a social botnet: Growth, content and influence in twitter.
\newblock In {\em Proceedings of the 18th ACM Conference on Computer Supported
  Cooperative Work \&\#38; Social Computing}, CSCW '15, Pp.~ 839--851, New
  York, NY, USA. ACM.

\bibitem[\protect\astroncite{Adewole et~al.}{2017}]{Adewole2017}
Adewole, K.~S., N.~B. Anuar, A.~Kamsin, K.~D. Varathan, and S.~A.
  Razak\leavevmode\nopagebreak\newline 2017.
\newblock Malicious accounts: dark of the social networks.
\newblock {\em Journal of Network and Computer Applications}, 79:41--67.

\bibitem[\protect\astroncite{Agarwal et~al.}{2011}]{agarwal2011sentiment}
Agarwal, A., B.~Xie, I.~Vovsha, O.~Rambow, and
  R.~Passonneau\leavevmode\nopagebreak\newline 2011.
\newblock Sentiment analysis of twitter data.
\newblock In {\em Proceedings of the Workshop on Language in Social Media (LSM
  2011)}, Pp.~ 30--38.

\bibitem[\protect\astroncite{Ahmed and Abulaish}{2013}]{ahmed2013generic}
Ahmed, F. and M.~Abulaish\leavevmode\nopagebreak\newline 2013.
\newblock A generic statistical approach for spam detection in online social
  networks.
\newblock {\em Computer Communications}, 36(10-11):1120--1129.

\bibitem[\protect\astroncite{Aiken and Kim}{2018}]{aiken2018poster}
Aiken, W. and H.~Kim\leavevmode\nopagebreak\newline 2018.
\newblock Poster: Deepcrack: Using deep learning to automatically crack audio
  captchas.
\newblock In {\em Proceedings of the 2018 on Asia Conference on Computer and
  Communications Security}, Pp.~ 797--799. ACM.

\bibitem[\protect\astroncite{Al-Qurishi et~al.}{2018a}]{al2018prediction}
Al-Qurishi, M., M.~Alrubaian, S.~M.~M. Rahman, A.~Alamri, and M.~M.
  Hassan\leavevmode\nopagebreak\newline 2018a.
\newblock A prediction system of sybil attack in social network using
  deep-regression model.
\newblock {\em Future Generation Computer Systems}, 87:743--753.

\bibitem[\protect\astroncite{Al-Qurishi et~al.}{2018b}]{al2018leveraging}
Al-Qurishi, M., M.~S. Hossain, M.~Alrubaian, S.~M.~M. Rahman, and
  A.~Alamri\leavevmode\nopagebreak\newline 2018b.
\newblock Leveraging analysis of user behavior to identify malicious activities
  in large-scale social networks.
\newblock {\em IEEE Transactions on Industrial Informatics}, 14(2):799--813.

\bibitem[\protect\astroncite{Alvisi et~al.}{2013}]{alvisi2013sok}
Alvisi, L., A.~Clement, A.~Epasto, S.~Lattanzi, and
  A.~Panconesi\leavevmode\nopagebreak\newline 2013.
\newblock Sok: The evolution of sybil defense via social networks.
\newblock In {\em Security and Privacy (SP), 2013 IEEE Symposium on}, Pp.~
  382--396. IEEE.

\bibitem[\protect\astroncite{Bakshy et~al.}{2011}]{Bakshy:2011}
Bakshy, E., J.~M. Hofman, W.~A. Mason, and D.~J.
  Watts\leavevmode\nopagebreak\newline 2011.
\newblock Everyone's an influencer: quantifying influence on twitter.
\newblock In {\em Proceedings of the fourth ACM international conference on Web
  search and data mining}, Pp.~ 65--74. ACM.

\bibitem[\protect\astroncite{Baltazar et~al.}{2009}]{Baltazar:2009}
Baltazar, J., J.~Costoya, and R.~Flores\leavevmode\nopagebreak\newline 2009.
\newblock The real face of koobface: The largest web 2.0 botnet explained.
\newblock {\em Trend Micro Research}, 5(9):10.

\bibitem[\protect\astroncite{Barbosa and Feng}{2010}]{barbosa2010robust}
Barbosa, L. and J.~Feng\leavevmode\nopagebreak\newline 2010.
\newblock Robust sentiment detection on twitter from biased and noisy data.
\newblock In {\em Proceedings of the 23rd international conference on
  computational linguistics: posters}, Pp.~ 36--44. Association for
  Computational Linguistics.

\bibitem[\protect\astroncite{Be{\u{g}}enilmi{\c{s}} and
  Uskudarli}{2018}]{Beugenilmics:2018}
Be{\u{g}}enilmi{\c{s}}, E. and S.~Uskudarli\leavevmode\nopagebreak\newline
  2018.
\newblock Organized behavior classification of tweet sets using supervised
  learning methods.
\newblock In {\em Proceedings of the 8th International Conference on Web
  Intelligence, Mining and Semantics}, P.~~36. ACM.

\bibitem[\protect\astroncite{Benamara et~al.}{2007}]{benamara2007sentiment}
Benamara, F., C.~Cesarano, A.~Picariello, D.~R. Recupero, and V.~S.
  Subrahmanian\leavevmode\nopagebreak\newline 2007.
\newblock Sentiment analysis: Adjectives and adverbs are better than adjectives
  alone.
\newblock In {\em ICWSM}, Pp.~ 1--7. Citeseer.

\bibitem[\protect\astroncite{Benevenuto et~al.}{2010}]{benevenuto2010detecting}
Benevenuto, F., G.~Magno, T.~Rodrigues, and
  V.~Almeida\leavevmode\nopagebreak\newline 2010.
\newblock Detecting spammers on twitter.
\newblock In {\em Collaboration, electronic messaging, anti-abuse and spam
  conference (CEAS)}, volume~6, P.~~12.

\bibitem[\protect\astroncite{Benigni et~al.}{2019}]{Benigni:2019}
Benigni, M.~C., K.~Joseph, and K.~M. Carley\leavevmode\nopagebreak\newline
  2019.
\newblock Bot-ivistm: Assessing information manipulation in social media using
  network analytics.
\newblock In {\em Emerging Research Challenges and Opportunities in
  Computational Social Network Analysis and Mining}, Pp.~ 19--42.
\newblock Springer.

\bibitem[\protect\astroncite{Besel et~al.}{2018}]{Besel:2018}
Besel, C., J.~Echeverria, and S.~Zhou\leavevmode\nopagebreak\newline 2018.
\newblock Full cycle analysis of a large-scale botnet attack on twitter.
\newblock In {\em 2018 IEEE/ACM International Conference on Advances in Social
  Networks Analysis and Mining (ASONAM)}, Pp.~ 170--177. IEEE.

\bibitem[\protect\astroncite{Bessi and Ferrara}{2016}]{FM7090}
Bessi, A. and E.~Ferrara\leavevmode\nopagebreak\newline 2016.
\newblock Social bots distort the 2016 u.s. presidential election online
  discussion.
\newblock {\em First Monday}, 21(11).

\bibitem[\protect\astroncite{Beutel et~al.}{2013}]{Copycatch}
Beutel, A., W.~Xu, V.~Guruswami, C.~Palow, and
  C.~Faloutsos\leavevmode\nopagebreak\newline 2013.
\newblock Copycatch: stopping group attacks by spotting lockstep behavior in
  social networks.
\newblock In {\em Proceedings of the 22nd international conference on World
  Wide Web}, Pp.~ 119--130. ACM.

\bibitem[\protect\astroncite{Blondel et~al.}{2008}]{blondel2008fast}
Blondel, V.~D., J.-L. Guillaume, R.~Lambiotte, and
  E.~Lefebvre\leavevmode\nopagebreak\newline 2008.
\newblock Fast unfolding of communities in large networks.
\newblock {\em Journal of statistical mechanics: theory and experiment},
  2008(10):P10008.

\bibitem[\protect\astroncite{Bnaya et~al.}{2013a}]{bnaya2013bandit}
Bnaya, Z., R.~Puzis, R.~Stern, and A.~Felner\leavevmode\nopagebreak\newline
  2013a.
\newblock Bandit algorithms for social network queries.
\newblock In {\em 2013 International Conference on Social Computing}, Pp.~
  148--153. IEEE.

\bibitem[\protect\astroncite{Bnaya et~al.}{2013b}]{bnaya2013social}
Bnaya, Z., R.~Puzis, R.~Stern, and A.~Felner\leavevmode\nopagebreak\newline
  2013b.
\newblock Social network search as a volatile multi-armed bandit problem.
\newblock {\em HUMAN}, 2(2):pp--84.

\bibitem[\protect\astroncite{Bock et~al.}{2017}]{bock2017uncaptcha}
Bock, K., D.~Patel, G.~Hughey, and D.~Levin\leavevmode\nopagebreak\newline
  2017.
\newblock uncaptcha: a low-resource defeat of recaptcha's audio challenge.
\newblock In {\em Proceedings of the 11th USENIX Conference on Offensive
  Technologies}, Pp.~ 7--7. USENIX Association.

\bibitem[\protect\astroncite{Boshmaf et~al.}{2015}]{boshmaf2015integro}
Boshmaf, Y., D.~Logothetis, G.~Siganos, J.~Ler{\'\i}a, J.~Lorenzo, M.~Ripeanu,
  and K.~Beznosov\leavevmode\nopagebreak\newline 2015.
\newblock Integro: Leveraging victim prediction for robust fake account
  detection in osns.
\newblock In {\em NDSS}, volume~15, Pp.~ 8--11.

\bibitem[\protect\astroncite{Boshmaf et~al.}{2011}]{boshmaf2011socialbot}
Boshmaf, Y., I.~Muslukhov, K.~Beznosov, and
  M.~Ripeanu\leavevmode\nopagebreak\newline 2011.
\newblock The socialbot network: when bots socialize for fame and money.
\newblock In {\em Proceedings of the 27th annual computer security applications
  conference}, Pp.~ 93--102. ACM.

\bibitem[\protect\astroncite{Boshmaf et~al.}{2013a}]{Boshmaf:2013}
Boshmaf, Y., I.~Muslukhov, K.~Beznosov, and
  M.~Ripeanu\leavevmode\nopagebreak\newline 2013a.
\newblock Design and analysis of a social botnet.
\newblock {\em Comput. Netw.}, 57(2):556--578.

\bibitem[\protect\astroncite{Boshmaf et~al.}{2013b}]{boshmaf2013design}
Boshmaf, Y., I.~Muslukhov, K.~Beznosov, and
  M.~Ripeanu\leavevmode\nopagebreak\newline 2013b.
\newblock Design and analysis of a social botnet.
\newblock {\em Computer Networks}, 57(2):556--578.

\bibitem[\protect\astroncite{Bruns and Stieglitz}{2014a}]{bruns2014metrics}
Bruns, A. and S.~Stieglitz\leavevmode\nopagebreak\newline 2014a.
\newblock Metrics for understanding communication on twitter.
\newblock In {\em Twitter and society}, volume~89, Pp.~ 69--82.
\newblock Peter Lang.

\bibitem[\protect\astroncite{Bruns and Stieglitz}{2014b}]{bruns2014}
Bruns, A. and S.~Stieglitz\leavevmode\nopagebreak\newline 2014b.
\newblock Twitter data: what do they represent?
\newblock {\em it-Information Technology}, 56(5):240--245.

\bibitem[\protect\astroncite{Burges et~al.}{2005}]{RankNet}
Burges, C., T.~Shaked, E.~Renshaw, A.~Lazier, M.~Deeds, N.~Hamilton, and G.~N.
  Hullender\leavevmode\nopagebreak\newline 2005.
\newblock Learning to rank using gradient descent.
\newblock In {\em Proceedings of the 22nd International Conference on Machine
  learning (ICML-05)}, Pp.~ 89--96.

\bibitem[\protect\astroncite{Burghouwt et~al.}{2013}]{Burghouwt:2013}
Burghouwt, P., M.~Spruit, and H.~Sips\leavevmode\nopagebreak\newline 2013.
\newblock Detection of covert botnet command and control channels by causal
  analysis of traffic flows.
\newblock In {\em Cyberspace Safety and Security}, Pp.~ 117--131.
\newblock Springer.

\bibitem[\protect\astroncite{Cai and Jermaine}{2012}]{cai2012latent}
Cai, Z. and C.~Jermaine\leavevmode\nopagebreak\newline 2012.
\newblock The latent community model for detecting sybil attacks in social
  networks.
\newblock In {\em Proc. NDSS}.

\bibitem[\protect\astroncite{Cao and Qiu}{2013}]{Cao:2013asp2p}
Cao, L. and X.~Qiu\leavevmode\nopagebreak\newline 2013.
\newblock Asp2p: An advanced botnet based on social networks over hybrid p2p.
\newblock In {\em Wireless and Optical Communication Conference (WOCC), 2013
  22nd}, Pp.~ 677--682. IEEE.

\bibitem[\protect\astroncite{Cao et~al.}{2012}]{cao2012aiding}
Cao, Q., M.~Sirivianos, X.~Yang, and
  T.~Pregueiro\leavevmode\nopagebreak\newline 2012.
\newblock Aiding the detection of fake accounts in large scale social online
  services.
\newblock In {\em Proceedings of the 9th USENIX conference on Networked Systems
  Design and Implementation}, Pp.~ 15--15. USENIX Association.

\bibitem[\protect\astroncite{Cao and Yang}{2013}]{cao2013sybilfence}
Cao, Q. and X.~Yang\leavevmode\nopagebreak\newline 2013.
\newblock Sybilfence: Improving social-graph-based sybil defenses with user
  negative feedback.
\newblock {\em arXiv preprint arXiv:1304.3819}.

\bibitem[\protect\astroncite{Cao et~al.}{2014}]{cao2014uncovering}
Cao, Q., X.~Yang, J.~Yu, and C.~Palow\leavevmode\nopagebreak\newline 2014.
\newblock Uncovering large groups of active malicious accounts in online social
  networks.
\newblock In {\em Proceedings of the 2014 ACM SIGSAC Conference on Computer and
  Communications Security}, Pp.~ 477--488. ACM.

\bibitem[\protect\astroncite{Chavoshi et~al.}{2016a}]{chavoshi2016debot}
Chavoshi, N., H.~Hamooni, and A.~Mueen\leavevmode\nopagebreak\newline 2016a.
\newblock Debot: Twitter bot detection via warped correlation.
\newblock In {\em ICDM}, Pp.~ 817--822.

\bibitem[\protect\astroncite{Chavoshi et~al.}{2016b}]{chavoshi2016identifying}
Chavoshi, N., H.~Hamooni, and A.~Mueen\leavevmode\nopagebreak\newline 2016b.
\newblock Identifying correlated bots in twitter.
\newblock In {\em International Conference on Social Informatics}, Pp.~ 14--21.
  Springer.

\bibitem[\protect\astroncite{Chen and Subramanian}{2018}]{Chen:2018}
Chen, Z. and D.~Subramanian\leavevmode\nopagebreak\newline 2018.
\newblock An unsupervised approach to detect spam campaigns that use botnets on
  twitter.
\newblock {\em arXiv preprint arXiv:1804.05232}.

\bibitem[\protect\astroncite{Chu et~al.}{2012}]{Chu:2012}
Chu, Z., S.~Gianvecchio, H.~Wang, and S.~Jajodia\leavevmode\nopagebreak\newline
  2012.
\newblock Detecting automation of twitter accounts: Are you a human, bot, or
  cyborg?
\newblock {\em IEEE Transactions on Dependable and Secure Computing},
  9(6):811--824.

\bibitem[\protect\astroncite{Clark et~al.}{2016}]{clark2016sifting}
Clark, E.~M., J.~R. Williams, C.~A. Jones, R.~A. Galbraith, C.~M. Danforth, and
  P.~S. Dodds\leavevmode\nopagebreak\newline 2016.
\newblock Sifting robotic from organic text: a natural language approach for
  detecting automation on twitter.
\newblock {\em Journal of computational science}, 16:1--7.

\bibitem[\protect\astroncite{Clauset et~al.}{2004}]{clauset2004finding}
Clauset, A., M.~E. Newman, and C.~Moore\leavevmode\nopagebreak\newline 2004.
\newblock Finding community structure in very large networks.
\newblock {\em Physical review E}, 70(6):066111.

\bibitem[\protect\astroncite{Clauset et~al.}{2009}]{clauset2009power}
Clauset, A., C.~R. Shalizi, and M.~E. Newman\leavevmode\nopagebreak\newline
  2009.
\newblock Power-law distributions in empirical data.
\newblock {\em SIAM review}, 51(4):661--703.

\bibitem[\protect\astroncite{Compagno et~al.}{2015}]{Compagno:2015boten}
Compagno, A., M.~Conti, D.~Lain, G.~Lovisotto, and L.~V.
  Mancini\leavevmode\nopagebreak\newline 2015.
\newblock Boten elisa: A novel approach for botnet c\&c in online social
  networks.
\newblock In {\em Communications and Network Security (CNS), 2015 IEEE
  Conference on}, Pp.~ 74--82. IEEE.

\bibitem[\protect\astroncite{Cresci et~al.}{2015}]{cresci2015fame}
Cresci, S., R.~Di~Pietro, M.~Petrocchi, A.~Spognardi, and
  M.~Tesconi\leavevmode\nopagebreak\newline 2015.
\newblock Fame for sale: efficient detection of fake twitter followers.
\newblock {\em Decision Support Systems}, 80:56--71.

\bibitem[\protect\astroncite{Cresci et~al.}{2016}]{cresci2016dna}
Cresci, S., R.~Di~Pietro, M.~Petrocchi, A.~Spognardi, and
  M.~Tesconi\leavevmode\nopagebreak\newline 2016.
\newblock Dna-inspired online behavioral modeling and its application to
  spambot detection.
\newblock {\em IEEE Intelligent Systems}, 31(5):58--64.

\bibitem[\protect\astroncite{Cresci et~al.}{2017}]{Cresci2017}
Cresci, S., R.~Di~Pietro, M.~Petrocchi, A.~Spognardi, and
  M.~Tesconi\leavevmode\nopagebreak\newline 2017.
\newblock The paradigm-shift of social spambots: Evidence, theories, and tools
  for the arms race.
\newblock In {\em Proceedings of the 26th International Conference on World
  Wide Web Companion}, Pp.~ 963--972. International World Wide Web Conferences
  Steering Committee.

\bibitem[\protect\astroncite{Cresci et~al.}{2018a}]{cresci2018social}
Cresci, S., R.~Di~Pietro, M.~Petrocchi, A.~Spognardi, and
  M.~Tesconi\leavevmode\nopagebreak\newline 2018a.
\newblock Social fingerprinting: detection of spambot groups through
  dna-inspired behavioral modeling.
\newblock {\em IEEE Transactions on Dependable and Secure Computing},
  15(4):561--576.

\bibitem[\protect\astroncite{Cresci et~al.}{2018b}]{Cresci:2018}
Cresci, S., F.~Lillo, D.~Regoli, S.~Tardelli, and
  M.~Tesconi\leavevmode\nopagebreak\newline 2018b.
\newblock Cashtag piggybacking: uncovering spam and bot activity in stock
  microblogs on twitter.
\newblock {\em arXiv preprint arXiv:1804.04406}.

\bibitem[\protect\astroncite{Cresci et~al.}{2018c}]{cresci2018reaction}
Cresci, S., M.~Petrocchi, A.~Spognardi, and
  S.~Tognazzi\leavevmode\nopagebreak\newline 2018c.
\newblock From reaction to proaction: Unexplored ways to the detection of
  evolving spambots.
\newblock In {\em Companion of the The Web Conference 2018 on The Web
  Conference 2018}, Pp.~ 1469--1470. International World Wide Web Conferences
  Steering Committee.

\bibitem[\protect\astroncite{Cresci et~al.}{2019}]{cresci2019capability}
Cresci, S., M.~Petrocchi, A.~Spognardi, and
  S.~Tognazzi\leavevmode\nopagebreak\newline 2019.
\newblock On the capability of evolved spambots to evade detection via genetic
  engineering.
\newblock {\em Online Social Networks and Media}, 9:1--16.

\bibitem[\protect\astroncite{Danezis and Mittal}{2009}]{danezis2009sybilinfer}
Danezis, G. and P.~Mittal\leavevmode\nopagebreak\newline 2009.
\newblock Sybilinfer: Detecting sybil nodes using social networks.
\newblock In {\em NDSS}, Pp.~ 1--15. San Diego, CA.

\bibitem[\protect\astroncite{Davis et~al.}{2016}]{Davis:2016}
Davis, C.~A., O.~Varol, E.~Ferrara, A.~Flammini, and
  F.~Menczer\leavevmode\nopagebreak\newline 2016.
\newblock Botornot: A system to evaluate social bots.
\newblock In {\em Proceedings of the 25th International Conference Companion on
  World Wide Web}, Pp.~ 273--274. International World Wide Web Conferences
  Steering Committee.

\bibitem[\protect\astroncite{De~Cristofaro et~al.}{2014}]{de2014paying}
De~Cristofaro, E., A.~Friedman, G.~Jourjon, M.~A. Kaafar, and M.~Z.
  Shafiq\leavevmode\nopagebreak\newline 2014.
\newblock Paying for likes?: Understanding facebook like fraud using honeypots.
\newblock In {\em Proceedings of the 2014 Conference on Internet Measurement
  Conference}, Pp.~ 129--136. ACM.

\bibitem[\protect\astroncite{Dickerson et~al.}{2014}]{Dickerson:2014}
Dickerson, J.~P., V.~Kagan, and V.~S.
  Subrahmanian\leavevmode\nopagebreak\newline 2014.
\newblock Using sentiment to detect bots on twitter: Are humans more
  opinionated than bots?
\newblock In {\em Proceedings of the 2014 IEEE/ACM International Conference on
  Advances in Social Networks Analysis and Mining}, ASONAM '14, Pp.~ 620--627,
  Piscataway, NJ, USA. IEEE Press.

\bibitem[\protect\astroncite{Dorri et~al.}{2018}]{Dorri:2018}
Dorri, A., M.~Abadi, and M.~Dadfarnia\leavevmode\nopagebreak\newline 2018.
\newblock Socialbothunter: Botnet detection in twitter-like social networking
  services using semi-supervised collective classification.
\newblock In {\em 2018 IEEE 16th Intl Conf on Dependable, Autonomic and Secure
  Computing, 16th Intl Conf on Pervasive Intelligence and Computing, 4th Intl
  Conf on Big Data Intelligence and Computing and Cyber Science and Technology
  Congress (DASC/PiCom/DataCom/CyberSciTech)}, Pp.~ 496--503. IEEE.

\bibitem[\protect\astroncite{Echeverr{\'\i}a et~al.}{2018}]{Echeverria:2018}
Echeverr{\'\i}a, J., E.~De~Cristofaro, N.~Kourtellis, I.~Leontiadis,
  G.~Stringhini, and S.~Zhou\leavevmode\nopagebreak\newline 2018.
\newblock Lobo--evaluation of generalization deficiencies in twitter bot
  classifiers.
\newblock {\em arXiv preprint arXiv:1809.09684}.

\bibitem[\protect\astroncite{Echeverria and Zhou}{2017a}]{Echeverria:2017burst}
Echeverria, J. and S.~Zhou\leavevmode\nopagebreak\newline 2017a.
\newblock Discovery of the twitter bursty botnet.
\newblock {\em arXiv preprint arXiv:1709.06740}.

\bibitem[\protect\astroncite{Gürsun et~al.}{2018}]{GoncaHoca}
Gürsun, G., M.~Sensoy, and M.~Kandemir\leavevmode\nopagebreak\newline 2018.
\newblock On context-aware ddos attacks using deep generative networks.
\newblock In {\em 2018 27th International Conference on Computer Communication
  and Networks (ICCCN)}, Pp.~ 1--9.
  
\bibitem[\protect\astroncite{Echeverria and Zhou}{2017b}]{Echeverria:2017star}
Echeverria, J. and S.~Zhou\leavevmode\nopagebreak\newline 2017b.
\newblock Discovery, retrieval, and analysis of the'star wars' botnet in
  twitter.
\newblock In {\em Proceedings of the 2017 IEEE/ACM International Conference on
  Advances in Social Networks Analysis and Mining 2017}, Pp.~ 1--8. ACM.

\bibitem[\protect\astroncite{El-Mawass et~al.}{2018}]{el2018supervised}
El-Mawass, N., P.~Honeine, and L.~Vercouter\leavevmode\nopagebreak\newline
  2018.
\newblock Supervised classification of social spammers using a similarity-based
  markov random field approach.
\newblock In {\em Proceedings of the 5th Multidisciplinary International Social
  Networks Conference}, P.~~14. ACM.

\bibitem[\protect\astroncite{Elishar et~al.}{2012}]{elishar2012organizational}
Elishar, A., M.~Fire, D.~Kagan, and Y.~Elovici\leavevmode\nopagebreak\newline
  2012.
\newblock Organizational intrusion: Organization mining using socialbots.
\newblock In {\em 2012 International Conference on Social Informatics}, Pp.~
  7--12. IEEE.

\bibitem[\protect\astroncite{Elyashar et~al.}{2013}]{Elyashar}
Elyashar, A., M.~Fire, D.~Kagan, and Y.~Elovici\leavevmode\nopagebreak\newline
  2013.
\newblock Homing socialbots: Intrusion on a specific organization's employee
  using socialbots.
\newblock In {\em Proceedings of the 2013 IEEE/ACM International Conference on
  Advances in Social Networks Analysis and Mining}, ASONAM '13, Pp.~
  1358--1365, New York, NY, USA. ACM.

\bibitem[\protect\astroncite{Faghani and Nguyen}{2012}]{Faghani2012socell}
Faghani, M.~R. and U.~T. Nguyen\leavevmode\nopagebreak\newline 2012.
\newblock Socellbot: A new botnet design to infect smartphones via online
  social networking.
\newblock In {\em CCECE}, Pp.~ 1--5.

\bibitem[\protect\astroncite{Faghani and Nguyen}{2018}]{faghani2018mobile}
Faghani, M.~R. and U.~T. Nguyen\leavevmode\nopagebreak\newline 2018.
\newblock Mobile botnets meet social networks: design and analysis of a new
  type of botnet.
\newblock {\em International Journal of Information Security}, Pp.~ 1--27.

\bibitem[\protect\astroncite{Fazil and Abulaish}{2018}]{Fazil:2018}
Fazil, M. and M.~Abulaish\leavevmode\nopagebreak\newline 2018.
\newblock A hybrid approach for detecting automated spammers in twitter.
\newblock {\em IEEE Transactions on Information Forensics and Security},
  13(11):2707--2719.

\bibitem[\protect\astroncite{Feng et~al.}{2017}]{feng2017groupfound}
Feng, B., Q.~Li, X.~Pan, J.~Zhang, and D.~Guo\leavevmode\nopagebreak\newline
  2017.
\newblock Groupfound: An effective approach to detect suspicious accounts in
  online social networks.
\newblock {\em International Journal of Distributed Sensor Networks},
  13(7):1550147717722499.

\bibitem[\protect\astroncite{Fernquist et~al.}{2018}]{fernquist2018}
Fernquist, J., L.~Kaati, and R.~Schroeder\leavevmode\nopagebreak\newline 2018.
\newblock Political bots and the swedish general election.
\newblock In {\em 2018 IEEE International Conference on Intelligence and
  Security Informatics (ISI)}, Pp.~ 124--129. IEEE.

\bibitem[\protect\astroncite{Ferrara}{2018}]{Ferrara:2018}
Ferrara, E.\leavevmode\nopagebreak\newline 2018.
\newblock Measuring social spam and the effect of bots on information diffusion
  in social media.
\newblock In {\em Complex Spreading Phenomena in Social Systems}, Pp.~
  229--255.
\newblock Springer.

\bibitem[\protect\astroncite{Ferrara et~al.}{2016}]{Ferrara:2016}
Ferrara, E., O.~Varol, C.~Davis, F.~Menczer, and
  A.~Flammini\leavevmode\nopagebreak\newline 2016.
\newblock The rise of social bots.
\newblock {\em Commun. ACM}, 59(7):96--104.

\bibitem[\protect\astroncite{Fields}{2018}]{Fields:2018}
Fields, J.\leavevmode\nopagebreak\newline 2018.
\newblock Botnet campaign detection on twitter.
\newblock {\em arXiv preprint arXiv:1808.09839}.

\bibitem[\protect\astroncite{Fire and Puzis}{2016}]{fire2016organization}
Fire, M. and R.~Puzis\leavevmode\nopagebreak\newline 2016.
\newblock Organization mining using online social networks.
\newblock {\em Networks and Spatial Economics}, 16(2):545--578.

\bibitem[\protect\astroncite{Freitas et~al.}{2015}]{Freitas:2015}
Freitas, C., F.~Benevenuto, S.~Ghosh, and
  A.~Veloso\leavevmode\nopagebreak\newline 2015.
\newblock Reverse engineering socialbot infiltration strategies in twitter.
\newblock In {\em Proceedings of the 2015 IEEE/ACM International Conference on
  Advances in Social Networks Analysis and Mining 2015}, ASONAM '15, Pp.~
  25--32, New York, NY, USA. ACM.

\bibitem[\protect\astroncite{Gao et~al.}{2012}]{Gao2012towards}
Gao, H., Y.~Chen, K.~Lee, D.~Palsetia, and A.~N.
  Choudhary\leavevmode\nopagebreak\newline 2012.
\newblock Towards online spam filtering in social networks.
\newblock In {\em NDSS}, volume~12, Pp.~ 1--16.

\bibitem[\protect\astroncite{Gao et~al.}{2010}]{Gao2010}
Gao, H., J.~Hu, C.~Wilson, Z.~Li, Y.~Chen, and B.~Y.
  Zhao\leavevmode\nopagebreak\newline 2010.
\newblock Detecting and characterizing social spam campaigns.
\newblock In {\em Proceedings of the 10th ACM SIGCOMM conference on Internet
  measurement}, Pp.~ 35--47. ACM.

\bibitem[\protect\astroncite{Gao et~al.}{2015}]{gao2015sybilframe}
Gao, P., N.~Z. Gong, S.~Kulkarni, K.~Thomas, and
  P.~Mittal\leavevmode\nopagebreak\newline 2015.
\newblock Sybilframe: A defense-in-depth framework for structure-based sybil
  detection.
\newblock {\em arXiv preprint arXiv:1503.02985}.

\bibitem[\protect\astroncite{Gao et~al.}{2018}]{SybilFuse}
Gao, P., B.~Wang, N.~Z. Gong, S.~R. Kulkarni, K.~Thomas, and
  P.~Mittal\leavevmode\nopagebreak\newline 2018.
\newblock Sybilfuse: Combining local attributes with global structure to
  perform robust sybil detection.
\newblock In {\em 2018 IEEE Conference on Communications and Network Security
  (CNS)}, Pp.~ 1--9.

\bibitem[\protect\astroncite{Ghosh et~al.}{2012}]{Ghosh:2012}
Ghosh, S., B.~Viswanath, F.~Kooti, N.~K. Sharma, G.~Korlam, F.~Benevenuto,
  N.~Ganguly, and K.~P. Gummadi\leavevmode\nopagebreak\newline 2012.
\newblock Understanding and combating link farming in the twitter social
  network.
\newblock In {\em Proceedings of the 21st International Conference on World
  Wide Web}, WWW '12, Pp.~ 61--70, New York, NY, USA. ACM.

\bibitem[\protect\astroncite{Giatsoglou et~al.}{2015}]{giatsoglou2015nd}
Giatsoglou, M., D.~Chatzakou, N.~Shah, A.~Beutel, C.~Faloutsos, and
  A.~Vakali\leavevmode\nopagebreak\newline 2015.
\newblock Nd-sync: Detecting synchronized fraud activities.
\newblock In {\em Pacific-Asia Conference on Knowledge Discovery and Data
  Mining}, Pp.~ 201--214. Springer.

\bibitem[\protect\astroncite{Gilani et~al.}{2017}]{gilani2017f}
Gilani, Z., E.~Kochmar, and J.~Crowcroft\leavevmode\nopagebreak\newline 2017.
\newblock Classification of twitter accounts into automated agents and human
  users.
\newblock In {\em Proceedings of the 2017 IEEE/ACM International Conference on
  Advances in Social Networks Analysis and Mining 2017}, Pp.~ 489--496. ACM.

\bibitem[\protect\astroncite{Gong et~al.}{2014}]{gong2014sybilbelief}
Gong, N.~Z., M.~Frank, and P.~Mittal\leavevmode\nopagebreak\newline 2014.
\newblock Sybilbelief: A semi-supervised learning approach for structure-based
  sybil detection.
\newblock {\em IEEE Transactions on Information Forensics and Security},
  9(6):976--987.

  
\bibitem[\protect\astroncite{Gong et~al.}{2012}]{gong2012evolution}
Gong, N.~Z., W.~Xu, L.~Huang, P.~Mittal, E.~Stefanov, V.~Sekar, and
  D.~Song\leavevmode\nopagebreak\newline 2012.
\newblock Evolution of social-attribute networks: measurements, modeling, and
  implications using google+.
\newblock In {\em Proceedings of the 2012 Internet Measurement Conference},
  Pp.~ 131--144. ACM.

\bibitem[\protect\astroncite{Grier et~al.}{2010}]{Grier:2010}
Grier, C., K.~Thomas, V.~Paxson, and M.~Zhang\leavevmode\nopagebreak\newline
  2010.
\newblock @spam: The underground on 140 characters or less.
\newblock In {\em Proceedings of the 17th ACM Conference on Computer and
  Communications Security}, CCS '10, Pp.~ 27--37, New York, NY, USA. ACM.

\bibitem[\protect\astroncite{Grimme et~al.}{2018}]{grimme2018changing}
Grimme, C., D.~Assenmacher, and L.~Adam\leavevmode\nopagebreak\newline 2018.
\newblock Changing perspectives: Is it sufficient to detect social bots?
\newblock In {\em International Conference on Social Computing and Social
  Media}, Pp.~ 445--461. Springer.

\bibitem[\protect\astroncite{Gupta et~al.}{2019}]{gupta2019malreg}
Gupta, S., P.~Kumaraguru, and T.~Chakraborty\leavevmode\nopagebreak\newline
  2019.
\newblock Malreg: Detecting and analyzing malicious retweeter groups.
\newblock In {\em Proceedings of the ACM India Joint International Conference
  on Data Science and Management of Data}, Pp.~ 61--69. ACM.

\bibitem[\protect\astroncite{He et~al.}{2017}]{he2017understand}
He, Y., Q.~Li, J.~Cao, Y.~Ji, and D.~Guo\leavevmode\nopagebreak\newline 2017.
\newblock Understanding socialbot behavior on end hosts.
\newblock {\em International Journal of Distributed Sensor Networks},
  13(2):1550147717694170.

\bibitem[\protect\astroncite{He et~al.}{2016}]{He:2016}
He, Y., G.~Zhang, J.~Wu, and Q.~Li\leavevmode\nopagebreak\newline 2016.
\newblock Understanding a prospective approach to designing malicious social
  bots.
\newblock {\em Security and Communication Networks}, 9(13):2157--2172.

\bibitem[\protect\astroncite{Heydari et~al.}{2015}]{heydari2015detection}
Heydari, A., M.~ali Tavakoli, N.~Salim, and
  Z.~Heydari\leavevmode\nopagebreak\newline 2015.
\newblock Detection of review spam: A survey.
\newblock {\em Expert Systems with Applications}, 42(7):3634--3642.

\bibitem[\protect\astroncite{H{\"o}ner et~al.}{2017}]{honer2017minimizing}
H{\"o}ner, J., S.~Nakajima, A.~Bauer, K.-R. M{\"u}ller, and
  N.~G{\"o}rnitz\leavevmode\nopagebreak\newline 2017.
\newblock Minimizing trust leaks for robust sybil detection.
\newblock In {\em Proceedings of the 34th International Conference on Machine
  Learning-Volume 70}, Pp.~ 1520--1528. JMLR. org.

\bibitem[\protect\astroncite{Huber et~al.}{2010}]{huber2010}
Huber, M., M.~Mulazzani, and E.~Weippl\leavevmode\nopagebreak\newline 2010.
\newblock Who on earth is “mr. cypher”: automated friend injection attacks
  on social networking sites.
\newblock In {\em IFIP International Information Security Conference}, Pp.~
  80--89. Springer.

\bibitem[\protect\astroncite{Hwang et~al.}{2012}]{Hwang:2012}
Hwang, T., I.~Pearce, and M.~Nanis\leavevmode\nopagebreak\newline 2012.
\newblock Socialbots: Voices from the fronts.
\newblock {\em interactions}, 19(2):38--45.

\bibitem[\protect\astroncite{Ji et~al.}{2016}]{Ji:2016}
Ji, Y., Y.~He, X.~Jiang, J.~Cao, and Q.~Li\leavevmode\nopagebreak\newline 2016.
\newblock Combating the evasion mechanisms of social bots.
\newblock {\em computers \& security}, 58:230--249.

\bibitem[\protect\astroncite{Jiang et~al.}{2014}]{jiang2014catchsync}
Jiang, M., P.~Cui, A.~Beutel, C.~Faloutsos, and
  S.~Yang\leavevmode\nopagebreak\newline 2014.
\newblock Catchsync: catching synchronized behavior in large directed graphs.
\newblock In {\em Proceedings of the 20th ACM SIGKDD international conference
  on Knowledge discovery and data mining}, Pp.~ 941--950. ACM.

\bibitem[\protect\astroncite{Jiang et~al.}{2016a}]{jiang2016catching}
Jiang, M., P.~Cui, A.~Beutel, C.~Faloutsos, and
  S.~Yang\leavevmode\nopagebreak\newline 2016a.
\newblock Catching synchronized behaviors in large networks: A graph mining
  approach.
\newblock {\em ACM Transactions on Knowledge Discovery from Data (TKDD)},
  10(4):35.

\bibitem[\protect\astroncite{Jiang et~al.}{2016b}]{jiang2016inferring}
Jiang, M., P.~Cui, A.~Beutel, C.~Faloutsos, and
  S.~Yang\leavevmode\nopagebreak\newline 2016b.
\newblock Inferring lockstep behavior from connectivity pattern in large
  graphs.
\newblock {\em Knowledge and Information Systems}, 48(2):399--428.

\bibitem[\protect\astroncite{Kartaltepe et~al.}{2010}]{Kartaltepe:2010}
Kartaltepe, E.~J., J.~A. Morales, S.~Xu, and
  R.~Sandhu\leavevmode\nopagebreak\newline 2010.
\newblock Social network-based botnet command-and-control: emerging threats and
  countermeasures.
\newblock In {\em International Conference on Applied Cryptography and Network
  Security}, Pp.~ 511--528. Springer.

\bibitem[\protect\astroncite{Kaur and Singh}{2016}]{kaur2016survey}
Kaur, R. and S.~Singh\leavevmode\nopagebreak\newline 2016.
\newblock A survey of data mining and social network analysis based anomaly
  detection techniques.
\newblock {\em Egyptian Informatics Journal}, 17(2):199--216.

\bibitem[\protect\astroncite{Kayes and Iamnitchi}{2017}]{kayes2017privacy}
Kayes, I. and A.~Iamnitchi\leavevmode\nopagebreak\newline 2017.
\newblock Privacy and security in online social networks: A survey.
\newblock {\em Online Social Networks and Media}, 3:1--21.

\bibitem[\protect\astroncite{Kergl et~al.}{2014}]{Kergl:2014}
Kergl, D., R.~Roedler, and S.~Seeber\leavevmode\nopagebreak\newline 2014.
\newblock On the endogenesis of twitter's spritzer and gardenhose sample
  streams.
\newblock In {\em Proceedings of the 2014 IEEE/ACM International Conference on
  Advances in Social Networks Analysis and Mining}, Pp.~ 357--364. IEEE Press.

\bibitem[\protect\astroncite{Koll et~al.}{2017}]{koll2017thank}
Koll, D., M.~Schwarzmaier, J.~Li, X.-Y. Li, and
  X.~Fu\leavevmode\nopagebreak\newline 2017.
\newblock Thank you for being a friend: an attacker view on
  online-social-network-based sybil defenses.
\newblock In {\em Distributed Computing Systems Workshops (ICDCSW), 2017 IEEE
  37th International Conference on}, Pp.~ 157--162. IEEE.

\bibitem[\protect\astroncite{Kumar and Shah}{2018}]{kumar2018false}
Kumar, S. and N.~Shah\leavevmode\nopagebreak\newline 2018.
\newblock False information on web and social media: A survey.
\newblock {\em arXiv preprint arXiv:1804.08559}.

\bibitem[\protect\astroncite{Latah}{2019}]{mypaper1}
Latah, M.\leavevmode\nopagebreak\newline 2019.
\newblock Artificial intelligence enabled software-defined networking: a
  comprehensive overview.
\newblock {\em IET Networks}, 8:79--99(20).

\bibitem[\protect\astroncite{Lee et~al.}{2011a}]{lee2011content}
Lee, K., J.~Caverlee, Z.~Cheng, and D.~Z. Sui\leavevmode\nopagebreak\newline
  2011a.
\newblock Content-driven detection of campaigns in social media.
\newblock In {\em Proceedings of the 20th ACM international conference on
  Information and knowledge management}, Pp.~ 551--556. ACM.

\bibitem[\protect\astroncite{Lee et~al.}{2011b}]{Lee:2011}
Lee, K., B.~D. Eoff, and J.~Caverlee\leavevmode\nopagebreak\newline 2011b.
\newblock Seven months with the devils: a long-term study of content polluters
  on twitter.
\newblock In {\em In AAAI Int’l Conference on Weblogs and Social Media
  (ICWSM}.

\bibitem[\protect\astroncite{Lee et~al.}{2014}]{Lee:2014}
Lee, K., J.~Mahmud, J.~Chen, M.~Zhou, and
  J.~Nichols\leavevmode\nopagebreak\newline 2014.
\newblock Who will retweet this?: Automatically identifying and engaging
  strangers on twitter to spread information.
\newblock In {\em Proceedings of the 19th international conference on
  Intelligent User Interfaces}, Pp.~ 247--256. ACM.

\bibitem[\protect\astroncite{Lee and Kim}{2014}]{lee2014early}
Lee, S. and J.~Kim\leavevmode\nopagebreak\newline 2014.
\newblock Early filtering of ephemeral malicious accounts on twitter.
\newblock {\em Computer Communications}, 54:48--57.

\bibitem[\protect\astroncite{Lehtio}{2015}]{Lehtio:2015}
Lehtio, A.\leavevmode\nopagebreak\newline 2015.
\newblock C\&c-as-a-service: Abusing third-party web services as c\&c channels.
\newblock {\em Virus Bulletin}.

\bibitem[\protect\astroncite{Leskovec et~al.}{2008}]{leskovec2008s}
Leskovec, J., K.~J. Lang, A.~Dasgupta, and M.~W.
  Mahoney\leavevmode\nopagebreak\newline 2008.
\newblock Statistical properties of community structure in large social and
  information networks.
\newblock In {\em Proceedings of the 17th international conference on World
  Wide Web}, Pp.~ 695--704. ACM.

\bibitem[\protect\astroncite{Liu et~al.}{2015}]{liu2015exploiting}
Liu, C., P.~Gao, M.~Wright, and P.~Mittal\leavevmode\nopagebreak\newline 2015.
\newblock Exploiting temporal dynamics in sybil defenses.
\newblock In {\em Proceedings of the 22nd ACM SIGSAC Conference on Computer and
  Communications Security}, Pp.~ 805--816. ACM.

\bibitem[\protect\astroncite{Lu et~al.}{2011}]{Lu:2011advancedp2p}
Lu, T.-T., H.-Y. Liao, and M.-F. Chen\leavevmode\nopagebreak\newline 2011.
\newblock An advanced hybrid p2p botnet 2.0.
\newblock In {\em ICEIS (3)}, Pp.~ 273--276. Citeseer.

\bibitem[\protect\astroncite{Ma et~al.}{2014}]{ma2014sybil}
Ma, W., S.-Z. Hu, Q.~Dai, T.-T. Wang, and Y.-F.
  Huang\leavevmode\nopagebreak\newline 2014.
\newblock Sybil-resist: A new protocol for sybil attack defense in social
  network.
\newblock In {\em International Conference on Applications and Techniques in
  Information Security}, Pp.~ 219--230. Springer.

\bibitem[\protect\astroncite{Makkar et~al.}{2017}]{Makkar:2017sociobot}
Makkar, I.~K., F.~D. Troia, C.~A. Visaggio, T.~H. Austin, and
  M.~Stamp\leavevmode\nopagebreak\newline 2017.
\newblock Sociobot: a twitter-based botnet.
\newblock {\em International Journal of Security and Networks}, 12(1):1--12.

\bibitem[\protect\astroncite{Mei et~al.}{2017}]{Mei:2017}
Mei, B., Y.~Xiao, H.~Li, X.~Cheng, and Y.~Sun\leavevmode\nopagebreak\newline
  2017.
\newblock Inference attacks based on neural networks in social networks.
\newblock In {\em Proceedings of the Fifth ACM/IEEE Workshop on Hot Topics in
  Web Systems and Technologies}, HotWeb '17, Pp.~ 10:1--10:6, New York, NY,
  USA. ACM.

\bibitem[\protect\astroncite{Mesnards and Zaman}{2018}]{mesnards2018detecting}
Mesnards, N. G.~d. and T.~Zaman\leavevmode\nopagebreak\newline 2018.
\newblock Detecting influence campaigns in social networks using the ising
  model.
\newblock {\em arXiv preprint arXiv:1805.10244}.

\bibitem[\protect\astroncite{Messias et~al.}{2013}]{Messias:2013}
Messias, J., L.~Schmidt, R.~Oliveira, and
  F.~Benevenuto\leavevmode\nopagebreak\newline 2013.
\newblock You followed my bot! transforming robots into influential users in
  twitter.
\newblock {\em First Monday}, 18(7).

\bibitem[\protect\astroncite{Metzler and Croft}{2007}]{CoordinateAscent}
Metzler, D. and W.~B. Croft\leavevmode\nopagebreak\newline 2007.
\newblock Linear feature-based models for information retrieval.
\newblock {\em Information Retrieval}, 10(3):257--274.

\bibitem[\protect\astroncite{Miller et~al.}{2014}]{miller2014twitter}
Miller, Z., B.~Dickinson, W.~Deitrick, W.~Hu, and A.~H.
  Wang\leavevmode\nopagebreak\newline 2014.
\newblock Twitter spammer detection using data stream clustering.
\newblock {\em Information Sciences}, 260:64--73.

\bibitem[\protect\astroncite{Minnich et~al.}{2017}]{minnich2017}
Minnich, A., N.~Chavoshi, D.~Koutra, and
  A.~Mueen\leavevmode\nopagebreak\newline 2017.
\newblock Botwalk: Efficient adaptive exploration of twitter bot networks.
\newblock In {\em Proceedings of the 2017 IEEE/ACM International Conference on
  Advances in Social Networks Analysis and Mining 2017}, Pp.~ 467--474. ACM.

\bibitem[\protect\astroncite{Mislove et~al.}{2010}]{mislove2010you}
Mislove, A., B.~Viswanath, K.~P. Gummadi, and
  P.~Druschel\leavevmode\nopagebreak\newline 2010.
\newblock You are who you know: inferring user profiles in online social
  networks.
\newblock In {\em Proceedings of the third ACM international conference on Web
  search and data mining}, Pp.~ 251--260. ACM.

\bibitem[\protect\astroncite{Mohaisen et~al.}{2010}]{mohaisen2010measuring}
Mohaisen, A., A.~Yun, and Y.~Kim\leavevmode\nopagebreak\newline 2010.
\newblock Measuring the mixing time of social graphs.
\newblock In {\em Proceedings of the 10th ACM SIGCOMM conference on Internet
  measurement}, Pp.~ 383--389. ACM.

\bibitem[\protect\astroncite{Moonesinghe and Tan}{2008}]{outrank:2008}
Moonesinghe, H. and P.-N. Tan\leavevmode\nopagebreak\newline 2008.
\newblock Outrank: a graph-based outlier detection framework using random walk.
\newblock {\em International Journal on Artificial Intelligence Tools},
  17(01):19--36.

\bibitem[\protect\astroncite{Morstatter et~al.}{2016a}]{Morstatter:2016}
Morstatter, F., H.~Dani, J.~Sampson, and H.~Liu\leavevmode\nopagebreak\newline
  2016a.
\newblock Can one tamper with the sample api?: Toward neutralizing bias from
  spam and bot content.
\newblock In {\em Proceedings of the 25th International Conference Companion on
  World Wide Web}, Pp.~ 81--82. International World Wide Web Conferences
  Steering Committee.

\bibitem[\protect\astroncite{Morstatter et~al.}{2016b}]{morstatter2016new}
Morstatter, F., L.~Wu, T.~H. Nazer, K.~M. Carley, and
  H.~Liu\leavevmode\nopagebreak\newline 2016b.
\newblock A new approach to bot detection: striking the balance between
  precision and recall.
\newblock In {\em Proceedings of the 2016 IEEE/ACM International Conference on
  Advances in Social Networks Analysis and Mining}, Pp.~ 533--540. IEEE Press.

\bibitem[\protect\astroncite{Motoyama et~al.}{2010}]{motoyama2010re}
Motoyama, M., K.~Levchenko, C.~Kanich, D.~McCoy, G.~M. Voelker, and
  S.~Savage\leavevmode\nopagebreak\newline 2010.
\newblock Re: Captchas-understanding captcha-solving services in an economic
  context.
\newblock In {\em USENIX Security Symposium}, volume~10, P.~~3.

\bibitem[\protect\astroncite{Mulamba et~al.}{2016}]{mulamba2016ybilradar}
Mulamba, D., I.~Ray, and I.~Ray\leavevmode\nopagebreak\newline 2016.
\newblock Sybilradar: A graph-structure based framework for sybil detection in
  on-line social networks.
\newblock In {\em IFIP International Information Security and Privacy
  Conference}, Pp.~ 179--193. Springer.

\bibitem[\protect\astroncite{Nagaraja et~al.}{2011}]{Nagaraja2011stegobot}
Nagaraja, S., A.~Houmansadr, P.~Piyawongwisal, V.~Singh, P.~Agarwal, and
  N.~Borisov\leavevmode\nopagebreak\newline 2011.
\newblock Stegobot: a covert social network botnet.
\newblock In {\em International Workshop on Information Hiding}, Pp.~ 299--313.
  Springer.

\bibitem[\protect\astroncite{Nappa et~al.}{2010}]{nappa2010take}
Nappa, A., A.~Fattori, M.~Balduzzi, M.~Dell’Amico, and
  L.~Cavallaro\leavevmode\nopagebreak\newline 2010.
\newblock Take a deep breath: a stealthy, resilient and cost-effective botnet
  using skype.
\newblock In {\em International Conference on Detection of Intrusions and
  Malware, and Vulnerability Assessment}, Pp.~ 81--100. Springer.

\bibitem[\protect\astroncite{Newman and Girvan}{2004}]{newman2004finding}
Newman, M.~E. and M.~Girvan\leavevmode\nopagebreak\newline 2004.
\newblock Finding and evaluating community structure in networks.
\newblock {\em Physical review E}, 69(2):026113.

\bibitem[\protect\astroncite{Oentaryo et~al.}{2016}]{oentaryo:2016}
Oentaryo, R.~J., A.~Murdopo, P.~K. Prasetyo, and E.-P.
  Lim\leavevmode\nopagebreak\newline 2016.
\newblock On profiling bots in social media.
\newblock In {\em International Conference on Social Informatics}, Pp.~
  92--109. Springer.

\bibitem[\protect\astroncite{Pantic and Husain}{2015}]{Pantic:2015covert}
Pantic, N. and M.~I. Husain\leavevmode\nopagebreak\newline 2015.
\newblock Covert botnet command and control using twitter.
\newblock In {\em Proceedings of the 31st annual computer security applications
  conference}, Pp.~ 171--180. ACM.

\bibitem[\protect\astroncite{Paradise et~al.}{2014}]{paradise2014anti}
Paradise, A., R.~Puzis, and A.~Shabtai\leavevmode\nopagebreak\newline 2014.
\newblock Anti-reconnaissance tools: Detecting targeted socialbots.
\newblock {\em IEEE Internet Computing}, 18(5):11--19.

\bibitem[\protect\astroncite{Perez et~al.}{2011}]{perez2011spot}
Perez, C., M.~Lemercier, B.~Birregah, and
  A.~Corpel\leavevmode\nopagebreak\newline 2011.
\newblock Spot 1.0: Scoring suspicious profiles on twitter.
\newblock In {\em 2011 International Conference on Advances in Social Networks
  Analysis and Mining}, Pp.~ 377--381. IEEE.

\bibitem[\protect\astroncite{Perna and Tagarelli}{2018}]{Perna:2018}
Perna, D. and A.~Tagarelli\leavevmode\nopagebreak\newline 2018.
\newblock Learning to rank social bots.
\newblock In {\em Proceedings of the 29th on Hypertext and Social Media}, HT
  '18, Pp.~ 183--191, New York, NY, USA. ACM.

\bibitem[\protect\astroncite{Pramanik et~al.}{2015}]{Pramanik:2015}
Pramanik, S., M.~Danisch, Q.~Wang, and B.~Mitra\leavevmode\nopagebreak\newline
  2015.
\newblock An empirical approach towards an efficient “whom to mention?”
  twitter app.
\newblock In {\em Twitter for Research, 1st International \& Interdisciplinary
  Conference}.

\bibitem[\protect\astroncite{Prince}{2012}]{prince:2012Flashback}
Prince, B.\leavevmode\nopagebreak\newline 2012.
\newblock Flashback botnet updated to include twitter as c\&c.

\bibitem[\protect\astroncite{Ramalingam and
  Chinnaiah}{2018}]{ramalingam2018fake}
Ramalingam, D. and V.~Chinnaiah\leavevmode\nopagebreak\newline 2018.
\newblock Fake profile detection techniques in large-scale online social
  networks: A comprehensive review.
\newblock {\em Computers \& Electrical Engineering}, 65:165--177.

\bibitem[\protect\astroncite{Rapoport}{1953}]{rapoport1953}
Rapoport, A.\leavevmode\nopagebreak\newline 1953.
\newblock Spread of information through a population with socio-structural
  bias: I. assumption of transitivity.
\newblock {\em The bulletin of mathematical biophysics}, 15(4):523--533.

\bibitem[\protect\astroncite{Sebastian et~al.}{2014}]{Sebastian2014}
Sebastian, S., S.~Ayyappan, and P.~Vinod\leavevmode\nopagebreak\newline 2014.
\newblock Framework for design of graybot in social network.
\newblock In {\em Advances in Computing, Communications and Informatics
  (ICACCI, 2014 International Conference on}, Pp.~ 2331--2336. IEEE.

\bibitem[\protect\astroncite{Singh et~al.}{2013}]{Singh:2012}
Singh, A., A.~H. Toderici, K.~Ross, and M.~Stamp\leavevmode\nopagebreak\newline
  2013.
\newblock Social networking for botnet command and control.
\newblock {\em International Journal of Computer Network \& Information
  Security}, 5(6):11--17.

\bibitem[\protect\astroncite{Sivakorn et~al.}{2016a}]{sivakorn2016}
Sivakorn, S., I.~Polakis, and A.~D. Keromytis\leavevmode\nopagebreak\newline
  2016a.
\newblock I am robot:(deep) learning to break semantic image captchas.
\newblock In {\em Security and Privacy (EuroS\&P), 2016 IEEE European Symposium
  on}, Pp.~ 388--403. IEEE.

\bibitem[\protect\astroncite{Sivakorn et~al.}{2016b}]{sivakorn2016m}
Sivakorn, S., J.~Polakis, and A.~D. Keromytis\leavevmode\nopagebreak\newline
  2016b.
\newblock I’m not a human: Breaking the google recaptcha.
\newblock {\em Black Hat}.

\bibitem[\protect\astroncite{Song et~al.}{2015}]{song2015crowdtarget}
Song, J., S.~Lee, and J.~Kim\leavevmode\nopagebreak\newline 2015.
\newblock Crowdtarget: Target-based detection of crowdturfing in online social
  networks.
\newblock In {\em Proceedings of the 22nd ACM SIGSAC Conference on Computer and
  Communications Security}, Pp.~ 793--804. ACM.

\bibitem[\protect\astroncite{Sridharan et~al.}{2012}]{sridharan2012twitter}
Sridharan, V., V.~Shankar, and M.~Gupta\leavevmode\nopagebreak\newline 2012.
\newblock Twitter games: how successful spammers pick targets.
\newblock In {\em Proceedings of the 28th Annual Computer Security Applications
  Conference}, Pp.~ 389--398. ACM.

\bibitem[\protect\astroncite{Steiner}{2014}]{Steiner2014bots}
Steiner, T.\leavevmode\nopagebreak\newline 2014.
\newblock Bots vs. wikipedians, anons vs. logged-ins (redux): A global study of
  edit activity on wikipedia and wikidata.
\newblock In {\em Proceedings of The International Symposium on Open
  Collaboration}, P.~~25. ACM.

\bibitem[\protect\astroncite{Stringhini et~al.}{2012}]{stringhini2012poultry}
Stringhini, G., M.~Egele, C.~Kruegel, and
  G.~Vigna\leavevmode\nopagebreak\newline 2012.
\newblock Poultry markets: on the underground economy of twitter followers.
\newblock {\em ACM SIGCOMM Computer Communication Review}, 42(4):527--532.

\bibitem[\protect\astroncite{Stringhini et~al.}{2010}]{Stringhini:2010}
Stringhini, G., C.~Kruegel, and G.~Vigna\leavevmode\nopagebreak\newline 2010.
\newblock Detecting spammers on social networks.
\newblock In {\em Proceedings of the 26th Annual Computer Security Applications
  Conference}, ACSAC '10, Pp.~ 1--9, New York, NY, USA. ACM.

\bibitem[\protect\astroncite{Subrahmanian et~al.}{2016}]{Subrahmanian:2016}
Subrahmanian, V.~S., A.~Azaria, S.~Durst, V.~Kagan, A.~Galstyan, K.~Lerman,
  L.~Zhu, E.~Ferrara, A.~Flammini, and
  F.~Menczer\leavevmode\nopagebreak\newline 2016.
\newblock The darpa twitter bot challenge.
\newblock {\em Computer}, 49(6):38--46.

\bibitem[\protect\astroncite{Subrahmanian and
  Reforgiato}{2008}]{subrahmanian2008ava}
Subrahmanian, V.~S. and D.~Reforgiato\leavevmode\nopagebreak\newline 2008.
\newblock Ava: Adjective-verb-adverb combinations for sentiment analysis.
\newblock {\em IEEE Intelligent Systems}, 23(4):43--50.

\bibitem[\protect\astroncite{Suh et~al.}{2010}]{Suh:2010}
Suh, B., L.~Hong, P.~Pirolli, and E.~H. Chi\leavevmode\nopagebreak\newline
  2010.
\newblock Want to be retweeted? large scale analytics on factors impacting
  retweet in twitter network.
\newblock In {\em 2010 IEEE Second International Conference on Social
  Computing}, Pp.~ 177--184. IEEE.

\bibitem[\protect\astroncite{Tan et~al.}{2013}]{tan2013unik}
Tan, E., L.~Guo, S.~Chen, X.~Zhang, and Y.~Zhao\leavevmode\nopagebreak\newline
  2013.
\newblock Unik: Unsupervised social network spam detection.
\newblock In {\em Proceedings of the 22nd ACM international conference on
  Information \& Knowledge Management}, Pp.~ 479--488. ACM.

\bibitem[\protect\astroncite{Tanner et~al.}{2010}]{tanner2010koobface}
Tanner, B.~K., G.~Warner, H.~Stern, and
  S.~Olechowski\leavevmode\nopagebreak\newline 2010.
\newblock Koobface: The evolution of the social botnet.
\newblock In {\em eCrime Researchers Summit (eCrime), 2010}, Pp.~ 1--10. IEEE.

\bibitem[\protect\astroncite{Tavares and Faisal}{2013}]{tavares:2013}
Tavares, G. and A.~Faisal\leavevmode\nopagebreak\newline 2013.
\newblock Scaling-laws of human broadcast communication enable distinction
  between human, corporate and robot twitter users.
\newblock {\em PloS one}, 8(7):e65774.

\bibitem[\protect\astroncite{Varol et~al.}{2017a}]{Onur:2017}
Varol, O., E.~Ferrara, C.~Davis, F.~Menczer, and
  A.~Flammini\leavevmode\nopagebreak\newline 2017a.
\newblock Online human-bot interactions: Detection, estimation, and
  characterization.

\bibitem[\protect\astroncite{Varol et~al.}{2017b}]{varol2017early}
Varol, O., E.~Ferrara, F.~Menczer, and
  A.~Flammini\leavevmode\nopagebreak\newline 2017b.
\newblock Early detection of promoted campaigns on social media.
\newblock {\em EPJ Data Science}, 6(1):13.

\bibitem[\protect\astroncite{Viswanath et~al.}{2011}]{viswanath2011analysis}
Viswanath, B., A.~Post, K.~P. Gummadi, and
  A.~Mislove\leavevmode\nopagebreak\newline 2011.
\newblock An analysis of social network-based sybil defenses.
\newblock {\em ACM SIGCOMM Computer Communication Review}, 41(4):363--374.

\bibitem[\protect\astroncite{Wagner et~al.}{2012}]{Wanger:2012}
Wagner, C., S.~Mitter, C.~K{\"o}rner, and
  M.~Strohmaier\leavevmode\nopagebreak\newline 2012.
\newblock When social bots attack: Modeling susceptibility of users in online
  social networks.
\newblock In {\em In Proceedings of the 2nd Workshop on Making Sense of
  Microposts held in conjunction with the 21st World Wide Web Conference 2012},
  Pp.~ 41--48. .

\bibitem[\protect\astroncite{Wang}{2010}]{Wang:2010a}
Wang, A.~H.\leavevmode\nopagebreak\newline 2010.
\newblock Detecting spam bots in online social networking sites: A machine
  learning approach.
\newblock In {\em Data and Applications Security and Privacy XXIV}, S.~Foresti
  and S.~Jajodia, eds., Pp.~ 335--342, Berlin, Heidelberg. Springer Berlin
  Heidelberg.

\bibitem[\protect\astroncite{Wang et~al.}{2017}]{wang2017sybilscar}
Wang, B., L.~Zhang, and N.~Z. Gong\leavevmode\nopagebreak\newline 2017.
\newblock Sybilscar: Sybil detection in online social networks via local rule
  based propagation.
\newblock In {\em INFOCOM 2017-IEEE Conference on Computer Communications,
  IEEE}, Pp.~ 1--9. IEEE.

\bibitem[\protect\astroncite{Wang et~al.}{2018}]{SybilBlind}
Wang, B., L.~Zhang, and N.~Z. Gong\leavevmode\nopagebreak\newline 2018.
\newblock Sybilblind: Detecting fake users in online social networks without
  manual labels.
\newblock In {\em Research in Attacks, Intrusions, and Defenses}, M.~Bailey,
  T.~Holz, M.~Stamatogiannakis, and S.~Ioannidis, eds., Pp.~ 228--249, Cham.
  Springer International Publishing.

\bibitem[\protect\astroncite{Wu et~al.}{2018}]{Wu:2018slbot}
Wu, D., B.~Fang, J.~Yin, F.~Zhang, and X.~Cui\leavevmode\nopagebreak\newline
  2018.
\newblock Slbot: A serverless botnet based on service flux.
\newblock In {\em 2018 IEEE Third International Conference on Data Science in
  Cyberspace (DSC)}, Pp.~ 181--188. IEEE.

\bibitem[\protect\astroncite{Wu et~al.}{2010}]{LambdaMART}
Wu, Q., C.~J. Burges, K.~M. Svore, and J.~Gao\leavevmode\nopagebreak\newline
  2010.
\newblock Adapting boosting for information retrieval measures.
\newblock {\em Information Retrieval}, 13(3):254--270.

\bibitem[\protect\astroncite{Xiang et~al.}{2011}]{Xiang:2011andbot}
Xiang, C., F.~Binxing, Y.~Lihua, L.~Xiaoyi, and
  Z.~Tianning\leavevmode\nopagebreak\newline 2011.
\newblock Andbot: towards advanced mobile botnets.
\newblock In {\em Proceedings of the 4th USENIX conference on Large-scale
  exploits and emergent threats}, Pp.~ 11--11. USENIX Association.

\bibitem[\protect\astroncite{Xu and Li}{2007}]{AdaRank}
Xu, J. and H.~Li\leavevmode\nopagebreak\newline 2007.
\newblock Adarank: a boosting algorithm for information retrieval.
\newblock In {\em Proceedings of the 30th annual international ACM SIGIR
  conference on Research and development in information retrieval}, Pp.~
  391--398. ACM.

\bibitem[\protect\astroncite{Xue et~al.}{2013}]{xue2013votetrust}
Xue, J., Z.~Yang, X.~Yang, X.~Wang, L.~Chen, and
  Y.~Dai\leavevmode\nopagebreak\newline 2013.
\newblock Votetrust: Leveraging friend invitation graph to defend against
  social network sybils.
\newblock In {\em INFOCOM, 2013 Proceedings IEEE}, Pp.~ 2400--2408. IEEE.

\bibitem[\protect\astroncite{Yan}{2013}]{Yan:2013}
Yan, G.\leavevmode\nopagebreak\newline 2013.
\newblock Peri-watchdog: Hunting for hidden botnets in the periphery of online
  social networks.
\newblock {\em Computer Networks}, 57(2):540--555.

\bibitem[\protect\astroncite{Yang et~al.}{2013}]{yang2013empirical}
Yang, C., R.~Harkreader, and G.~Gu\leavevmode\nopagebreak\newline 2013.
\newblock Empirical evaluation and new design for fighting evolving twitter
  spammers.
\newblock {\em IEEE Transactions on Information Forensics and Security},
  8(8):1280--1293.

\bibitem[\protect\astroncite{Yang et~al.}{2011}]{yang2011free}
Yang, C., R.~C. Harkreader, and G.~Gu\leavevmode\nopagebreak\newline 2011.
\newblock Die free or live hard? empirical evaluation and new design for
  fighting evolving twitter spammers.
\newblock In {\em International Workshop on Recent Advances in Intrusion
  Detection}, Pp.~ 318--337. Springer.

\bibitem[\protect\astroncite{Yang et~al.}{2014}]{yang2014uncovering}
Yang, Z., C.~Wilson, X.~Wang, T.~Gao, B.~Y. Zhao, and
  Y.~Dai\leavevmode\nopagebreak\newline 2014.
\newblock Uncovering social network sybils in the wild.
\newblock {\em ACM Transactions on Knowledge Discovery from Data (TKDD)},
  8(1):2.

\bibitem[\protect\astroncite{Ye et~al.}{2018}]{Ye:2018:Captcha}
Ye, G., Z.~Tang, D.~Fang, Z.~Zhu, Y.~Feng, P.~Xu, X.~Chen, and
  Z.~Wang\leavevmode\nopagebreak\newline 2018.
\newblock Yet another text captcha solver: A generative adversarial network
  based approach.
\newblock In {\em Proceedings of the 2018 ACM SIGSAC Conference on Computer and
  Communications Security}, CCS '18, Pp.~ 332--348, New York, NY, USA. ACM.

\bibitem[\protect\astroncite{Yin et~al.}{2014}]{yin:2014drsn}
Yin, T., Y.~Zhang, and S.~Li\leavevmode\nopagebreak\newline 2014.
\newblock Dr-snbot: A social network-based botnet with strong
  destroy-resistance.
\newblock In {\em Networking, Architecture, and Storage (NAS), 2014 9th IEEE
  International Conference on}, Pp.~ 191--199. IEEE.

\bibitem[\protect\astroncite{Yu et~al.}{2008}]{yu2008sybillimit}
Yu, H., P.~B. Gibbons, M.~Kaminsky, and F.~Xiao\leavevmode\nopagebreak\newline
  2008.
\newblock Sybillimit: A near-optimal social network defense against sybil
  attacks.
\newblock In {\em 2008 IEEE Symposium on Security and Privacy (sp 2008)}, Pp.~
  3--17. IEEE.

\bibitem[\protect\astroncite{Yu et~al.}{2006}]{yu2006sybilguard}
Yu, H., M.~Kaminsky, P.~B. Gibbons, and
  A.~Flaxman\leavevmode\nopagebreak\newline 2006.
\newblock Sybilguard: defending against sybil attacks via social networks.
\newblock In {\em ACM SIGCOMM Computer Communication Review}, volume~36, Pp.~
  267--278. ACM.

\bibitem[\protect\astroncite{Zhang and Paxson}{2011}]{zhang2011detect}
Zhang, C.~M. and V.~Paxson\leavevmode\nopagebreak\newline 2011.
\newblock Detecting and analyzing automated activity on twitter.
\newblock In {\em International Conference on Passive and Active Network
  Measurement}, Pp.~ 102--111. Springer.

\bibitem[\protect\astroncite{Zhang et~al.}{2016a}]{zhang2016your}
Zhang, J., X.~Hu, Y.~Zhang, and H.~Liu\leavevmode\nopagebreak\newline 2016a.
\newblock Your age is no secret: Inferring microbloggers' ages via content and
  interaction analysis.
\newblock In {\em Tenth International AAAI Conference on Web and Social Media}.

\bibitem[\protect\astroncite{Zhang et~al.}{2013}]{zhang2013impact}
Zhang, J., R.~Zhang, Y.~Zhang, and G.~Yan\leavevmode\nopagebreak\newline 2013.
\newblock On the impact of social botnets for spam distribution and
  digital-influence manipulation.
\newblock In {\em Communications and Network Security (CNS), 2013 IEEE
  Conference on}, Pp.~ 46--54. IEEE.

\bibitem[\protect\astroncite{Zhang et~al.}{2016b}]{zhang2016rise}
Zhang, J., R.~Zhang, Y.~Zhang, and G.~Yan\leavevmode\nopagebreak\newline 2016b.
\newblock The rise of social botnets: Attacks and countermeasures.
\newblock {\em IEEE Transactions on Dependable and Secure Computing}.

\bibitem[\protect\astroncite{Zhang et~al.}{2018}]{zhang2018sybil}
Zhang, X., H.~Xie, and J.~C. Lui\leavevmode\nopagebreak\newline 2018.
\newblock Sybil detection in social-activity networks: Modeling, algorithms and
  evaluations.
\newblock In {\em 2018 IEEE 26th International Conference on Network Protocols
  (ICNP)}, Pp.~ 44--54. IEEE.

\bibitem[\protect\astroncite{Zhao et~al.}{2018}]{zhao2018actionable}
Zhao, T., M.~Malir, and M.~Jiang\leavevmode\nopagebreak\newline 2018.
\newblock Actionable objective optimization for suspicious behavior detection
  on large bipartite graphs.
\newblock In {\em 2018 IEEE International Conference on Big Data (Big Data)},
  Pp.~ 1248--1257. IEEE.

\end{thebibliography}
\end{document}